\documentclass[longnamesfirst]{emulateapj}
\usepackage{natbib}
\usepackage{apjfonts,lscape}
\tighten

\received{August 8, 2009}
\accepted{December 18, 2009}

\slugcomment{The Astrophysical Journal, in press}
\shortauthors{BAUER ET AL.}
\shorttitle{X-ray Constraints on AGN Properties in $z\sim2$ ULIRGS}

\begin{document}

\title{X-ray Constraints on the AGN Properties in {\it Spitzer}-IRS identified $\MakeLowercase{z}\sim2$  Ultraluminous Infrared Galaxies}

\author{
F.~E.~Bauer,\altaffilmark{1, 2, 3} Lin~Yan,\altaffilmark{4}
A.~Sajina,\altaffilmark{5} D.~M.~Alexander\altaffilmark{6} }

\altaffiltext{1}{Space Science Institute, 4750 Walnut Street, Suite 205, Boulder, Colorado 80301}
\altaffiltext{2}{Pontificia Universidad Cat\'{o}lica de Chile, Departamento de Astronom\'{\i}a y Astrof\'{\i}sica, Casilla 306, Santiago 22, Chile}
\altaffiltext{3}{Columbia Astrophysics Laboratory, 550 W. 120th St., Columbia University, New York, NY 10027}
\altaffiltext{4}{Spitzer Science Center, Caltech, MS 220-6, Pasadena, CA 91125}
\altaffiltext{5}{Haverford College, Haverford, PA 19041}
\altaffiltext{6}{Department of Physics, Durham University, Durham DH1 3LE, UK}

\begin{abstract}
We report {\it Chandra} \hbox{X-ray} constraints for 20 of the 52
high-redshift ultraluminous infrared galaxies (ULIRGs) identified in
the {\it Spitzer} Extragalactic First Look Survey with
$f_{\nu}(24\,\mu{\rm m})>0.9$~mJy, $\log ({\nu f_{\nu}(24\,\mu{\rm
m})\over \nu f_{\nu}(R)})>1$, and $\log ({\nu f_{\nu}(24\,\mu{\rm
m})\over \nu f_{\nu}(8\,\mu{\rm m})})>0.5$. Notably, decomposition of
{\it Spitzer} mid-infrared IRS spectra for the entire sample indicates
that they are comprised predominantly of weak-PAH ULIRGs dominated by
hot-dust continua, characteristic of AGN activity. Given their
redshifts, they have AGN bolometric luminosities
of \hbox{$\approx10^{45}$--$10^{47}$~erg~s$^{-1}$} comparable to
powerful Quasi-Stellar Objects (QSOs). This, coupled with their high
IR-to-optical ratios and often significant silicate absorption,
strongly argues in favor of these mid-IR objects being heavily
obscured QSOs. Here we use Chandra observations to further constrain
their obscuration.  At \hbox{X-ray} energies, we marginally detect two
ULIRGs, while the rest have only upper limits. Using the IRS-derived
5.8\,$\mu$m AGN continuum luminosity as a proxy for the
expected \hbox{X-ray} luminosities, we find that all of the observed
sources must individually be highly obscured, while \hbox{X-ray}
stacking limits on the undetected sources suggest that the majority,
if not all, are likely to be at least mildly
Compton-thick \hbox{($N_{\rm H}$$\ga$10$^{24}$~cm$^{-2}$)}. With a
space density of \hbox{$\approx$(1.4)$\times$10$^{-7}$~Mpc$^{-3}$} at
z$\sim$2, such objects imply an obscured AGN fraction (i.e., the ratio
of AGN above and below \hbox{$N_{\rm H}$$=$10$^{22}$~cm$^{-2}$})
of \hbox{$\ga$$1.7$:1} even among luminous QSOs. Given
that we do not correct for mid-IR extinction effects and that our
ULIRG selection is by no means complete for obscured AGN, we regard
our constraints as a lower limit to the true obscured fraction among
QSOs at this epoch.  Our findings, which are based on extensive
multi-wavelength constraints including {\it Spitzer} IRS spectra,
should aid in the interpretation of similar objects from larger or
deeper mid-IR surveys, where considerable uncertainty about the source
properties remains and comparable follow-up is not yet feasible.
\end{abstract}

\keywords{
galaxies: active
galaxies: high-redshift
infrared: galaxies
\hbox{X-ray}s: galaxies
}

\section{Introduction}\label{intro}
One of the most fundamental discoveries borne out of the past two
decades is that 1) local, massive galaxies almost universally host
central, super-massive black holes (SMBH; $\ga$ 10$^{6}$~M$_{\odot}$)
and 2) there exists a tight, fundamental correlation between a
galaxy's SMBH mass, stellar velocity dispersion, and bulge
luminosity \citep{Kormendy1995, Magorrian1998, Ferrarese2000}. This
key result implies that the evolution of a galaxy and its SMBH are
strongly coupled from a relatively early age. Indirect verification
comes from the broad agreement between the local SMBH mass density, as
inferred from the luminosity function of bulges (which itself should
be intimately related to galaxy evolution), and the integrated Active
Galactic Nuclei (AGN) emissivity estimated from measurements of the
Cosmic \hbox{X-ray} Background (CXRB) and the \hbox{X-ray} luminosity
function \citep[e.g.,][]{Yu2002, Marconi2004, Merloni2004,
LaFranca2005, Shankar2007}. However, direct observations of such
symbiotic growth have been difficult to obtain. A vital step is to
obtain an accurate census of all AGN activity at all epochs to truly
trace the growth of SMBHs.

Locally, highly obscured (i.e., Compton-thick with \hbox{$N_{\rm
H}$$\ga$10$^{24}$~cm$^{-2}$)} accretion appears to occur in $\sim$50\%
of identified AGN \citep[e.g.,][]{Risaliti1999,
Guainazzi2005, Malizia2009}. Similarly strong constraints at high redshift,
however, remain elusive. Notably, deep \hbox{X-ray} surveys have revealed an
AGN sky density of $\ga7200$~deg$^{-2}$, at least \hbox{2--10} times more
than are found at other wavelengths \citep[e.g.,][]{Bauer2004}. The
bulk of these faint AGN have redshifts of \hbox{0.5--2.0} and do indeed
appear to be obscured by significant gas and
dust \citetext{e.g., \citealp{Barger2003b};
\citealp{Szokoly2004}; see \citealp{Brandt2005} 
for a review}. Somewhat surprising, however, is the fact that such
sensitive \hbox{X-ray} surveys \citep{Alexander2003a, Luo2008} have yet to
uncover more than a handful of potential Compton-thick sources similar
to the ones found locally \citep[e.g.,][]{Bauer2004, Tozzi2006},
despite several tentative lines of evidence which suggest that
Compton-thick AGN could be at least as plentiful as found
locally \citep[e.g.,][]{Worsley2005, Treister2005, Alonso-Herrero2006,
Donley2007, Gilli2007, Fiore2009}. Thus constraints on this hidden
population of high-z Compton-thick AGN must rely on other means.

\begin{deluxetable*}{lrrrrrrrl}
\tabletypesize{\scriptsize}
\tablewidth{0pt}
\tablecaption{IR-Bright Candidate Compton-Thick ULIRGs\label{tab:brightsample}}
\tablehead{
\colhead{Object} & 
\colhead{z} & 
\colhead{$L^{\rm o}_{\rm 5.8\mu m}$} & 
\colhead{$L_{\rm 3-1000\mu m}$} & 
\colhead{$L^{\rm o}_{\rm 2-10\,keV}$} & 
\colhead{$L^{\rm oc}_{\rm 2-10\,keV}$} & 
\colhead{$N_{\rm H}$} & 
\colhead{$L^{\rm ic}_{\rm 2-10\,keV}$} & 
\colhead{References}
}
\tableheadfrac{0.05}
\startdata
\object{Mrk 231}            & 0.0422 &  45.10 &  46.14 & 42.46 & 42.46 & 24.3 & 43.65--44.30\rlap{*} & 1 \\
\object{Mrk 273}            & 0.0378 &  43.80 &  45.78 & 42.26 & 42.85 & 24.2 & 43.40\rlap{*} & 2, 3 \\
\object{Mrk 463}            & 0.0504 &  44.80 &  45.30 & 42.55 & 43.36 & 23.9 & 44.80         & 4 \\
\object{UGC 5101}           & 0.0394 &  44.10 &  45.59 & 42.34 & 42.84 & 24.1 & 44.30         & 5, 6 \\
\object{NGC 6240}           & 0.0245 &  43.50 &  45.44 & 42.09 & 42.54 & 24.3 & 44.20\rlap{*} & 7, 8, 9 \\
\object{IRAS 00182-7112}    & 0.3270 &  45.70 &  46.98 & 43.72 & 43.90 & 24.6 & 45.00         & 10 \\
\object{IRAS 09104+4109}    & 0.4420 &  46.10 &  46.82 & 44.15 & 44.15 & 24.5 & 46.10\rlap{*} & 11 \\
\object{IRAS 12514+1027}    & 0.3000 &  45.60 &  46.57 & 42.73 & 42.73 & $\ga$24.0 & 44.20         & 14 \\
\object{IRAS F15307+3252}   & 0.9257 &  46.00 &  47.26 & 43.78 & 43.78 & $\ga$24.0 & 45.30         & 12 \\
\object{IRAS 19254-7245}    & 0.0617 &  44.50 &  45.70 & 42.40 & 42.40 & 24.5 & 44.40\rlap{*} & 13 
\enddata
\tablecomments{
{\it Col. (1)} Object.
{\it Col. (2)} Redshift.
{\it Col. (3)} Logarithm of the rest-frame 5.8\,$\mu$m continuum
luminosity, calculated following S07, in units of erg~s$^{-1}$.
{\it Col. (4)} Logarithm of the rest-frame \hbox{3--1000,$\mu$m} continuum
luminosity ($L_{\rm IR}$), in units of erg~s$^{-1}$.
{\it Col. (5)} Logarithm of the observed rest-frame \hbox{2--10~keV}
luminosity of the AGN component derived directly from \hbox{2--10~keV} data
alone, with no correction for absorption, in units of
erg~s$^{-1}$. Since the \hbox{X-ray} spectra of every source is contaminated
by some degree of vigorous circumnuclear star formation, the values
quoted here should be considered approximate.
{\it Col. (6)} Logarithm of the rest-frame \hbox{2--10~keV} luminosity of
the AGN derived directly from spectral fits to \hbox{2--10~keV} data alone,
corrected for obvious absorption detected only in the \hbox{2--10~keV} band
itself, in units of erg~s$^{-1}$. In some cases, contamination from
vigorous circumnuclear star formation precluded any assessment of
apparent obscuration, in which case the values here are identical to
those in Col. (5).
{\it Col. (7)} Logarithm of the absorption column density toward the
AGN derived from spectral fitting to all available \hbox{X-ray} data, in
units of cm$^{-2}$.
{\it Col (8)} Logarithm of the rest-frame intrinsic \hbox{2--10~keV}
luminosity of the AGN derived from spectral fits to all available
\hbox{X-ray} data, assuming the \hbox{2--10~keV} data is due entirely to scattering
and/or reflection components, in units of erg~s$^{-1}$. Sources
denoted by '*' have been detected in the \hbox{10--40~keV} band, assumed to
be largely direct continuum, and should therefore be considered more
robust. Nonetheless, all absorption corrections are still model
dependent and should be considered approximate. Details regarding the
\hbox{X-ray} flux, measured column density, and absorption corrections can be
found in the references provided in Col. (9).
{\it Col. (9)} References. 1 --- \citealp{Braito2004}; 2 --- \citealp{Balestra2005}; 3 --- \citealp{Teng2009}; 4 --- \citealp{Bianchi2008}; 5 --- \citealp{Imanishi2003}; 6 --- \citealp{dellaCeca2008}; 7 --- \citealp{Vignati1999} 8 --- \citealp{Komossa2003}; 9 --- \citealp{Iwasawa2009}; 10 --- \citealp{Nandra2007}; 11 --- \citealp{Iwasawa2001}; 12 --- \citealp{Wilman2003}; 13 --- \citealp{Iwasawa2005}; 14 ---\citealp{Braito2009}.
}
\end{deluxetable*}

The mid-infrared (mid-IR) regime offers much potential for discovery,
since any primary AGN continuum that is absorbed must ultimately come
out at these wavelengths. Indeed, ultraluminous infrared galaxies
(ULIRGs, \hbox{$L_{\rm IR}$$\ga$10$^{12}$$L_{\odot}$}) and their
low-luminosity brethren have long stood out as candidates for highly
obscured accretion onto SMBHs \citep[e.g.,][]{Sanders1996, Farrah2003,
Farrah2007, Nardini2008}, and several of the closest and/or brightest
members of this class have now been confirmed as Compton-thick
AGN \citetext{e.g., \citealp{Comastri2004}; \citealp{dellaCeca2008};
see also Table~\ref{tab:brightsample}}. As a class, ULIRGs are almost
universally \hbox{X-ray} faint due to their obscured
nature \citep[e.g.,][]{Franceschini2003, Teng2005, Ruiz2007}, and have
relative AGN contributions which scale with increasing infrared
\hbox{(3--1000$\mu$m)} luminosity, often dominating the bolometric output in
the most luminous objects. This population has comparable space
densities to Quasi-stellar
Objects \citep[QSOs; e.g.,][]{Smail1997, Genzel2000} and has been
proposed as an important early evolutionary stage of these AGN
\citep[e.g.,][]{Page2004, Alexander2005a, Stevens2005}.
As such, ULIRGs could conceivably host a sizable fraction of the most
highly obscured, powerful AGN \hbox{($L_{\rm
bol}\ga10^{46}$~erg~s$^{-1}$)} that have evaded detection even in the
deepest \hbox{X-ray} surveys. The combination of mid-IR color
selection, which traces the hot dust that obscures AGN at most other
wavelengths \citep[e.g.,][]{Lacy2004, Stern2005}, and sensitive {\it
Spitzer} data, for instance, routinely offer efficient selection of
bright obscured and unobscured AGN by the hundreds now. Likewise,
simple 24$\mu$m flux selection may preferentially single out
AGN-dominated sources in the mid-IR, even when the bolometric emission
from such objects may still be
starburst-dominated \citep[e.g.,][]{Brand2006, Watabe2009}.

We focus here on the characterization of a sample of high-redshift,
{\it Spitzer}-selected \hbox{ULIRGs} which appear to host very
powerful obscured AGN based on their mid-IR spectra and UV-to-radio
spectral energy distributions (SEDs). Our goal is to understand how
these sources, which represent the extreme of the overall IR-emitting
galaxy population, fit into the picture of AGN demography. This paper
is organized as follows:
$\S$\ref{sample} describes our high-redshift sample;
$\S$\ref{data} details our data and reduction methods;
$\S$\ref{discussion} compares our new \hbox{X-ray} constraints to existing 
properties of IR-Bright ULIRGs, examines other AGN and star formation
indicators from provided by \citet[][hereafter S07 and S08,
respectfully]{Sajina2007, Sajina2008}, and investigates how these
sources relate to other selection techniques;
and finally $\S$\ref{conclude} summarizes our findings.
We adopt a flat $\Omega_{\Lambda},\Omega_{M}=0.7,0.3$ cosmology with
$h=H_{\rm 0}$(km~s$^{-1}$~Mpc$^{-1}$)$/100=0.70$, and a neutral
hydrogen column density of \hbox{$N_{\rm H}=(2.5\pm0.3)\times10^{20}$
cm$^{-2}$} for the $\approx$3.7~deg$^{-2}$ {\sl Spitzer} Extragalactic
First Look Survey
(xFLS)\footnote{\url{http://ssc.spitzer.caltech.edu/fls/}} region
\citep{Lockman2005}.

\section{Sample}\label{sample}
Our parent sample is comprised of 52 sources with $f_{\nu}(24\,\mu
{\rm m})>0.9$~mJy, $\log ({\nu f_{\nu}(24\,\mu{\rm m})\over \nu
f_{\nu}(R)})>1$, and $\log ({\nu f_{\nu}(24\,\mu{\rm m})\over \nu
f_{\nu}(8\,\mu {\rm m})})>0.5$ \citep{Yan2005}, selected from the full
$24\,\mu$m xFLS catalog. As noted in \citet[][hereafter Y07]{Yan2007},
this technique selects 59 objects over the 3.7 deg$^{-1}$ xFLS region
from the final mid-IR catalogs, of which 52 were originally
followed-up with {\it Spitzer} IRS using the initial mid-IR
catalogs. Although some objects technically lie just below the ULIRG
luminosity cutoff, for simplicity we hereafter refer to this sample as
xFLS ULIRGs. The two color criteria are designed to target $z\sim2$
IR-luminous obscured objects. As demonstrated in Fig.~2
of \citet{Yan2004}, typical starburst and obscured AGN spectral
templates migrate into this general color-color region by
\hbox{$z\sim1.5$--2}, while unobscured AGNs, by contrast, remain consistently
blue as a function of redshift, and hence are not selected. The
adopted 24\,$\mu$m flux density cutoff additionally limits our sample
to only the most luminous ULIRGs at $z\sim2$.

\begin{figure*}[t]
\vspace{0.0in}
\centerline{
\hglue-0.5cm{\includegraphics[width=17.0cm]{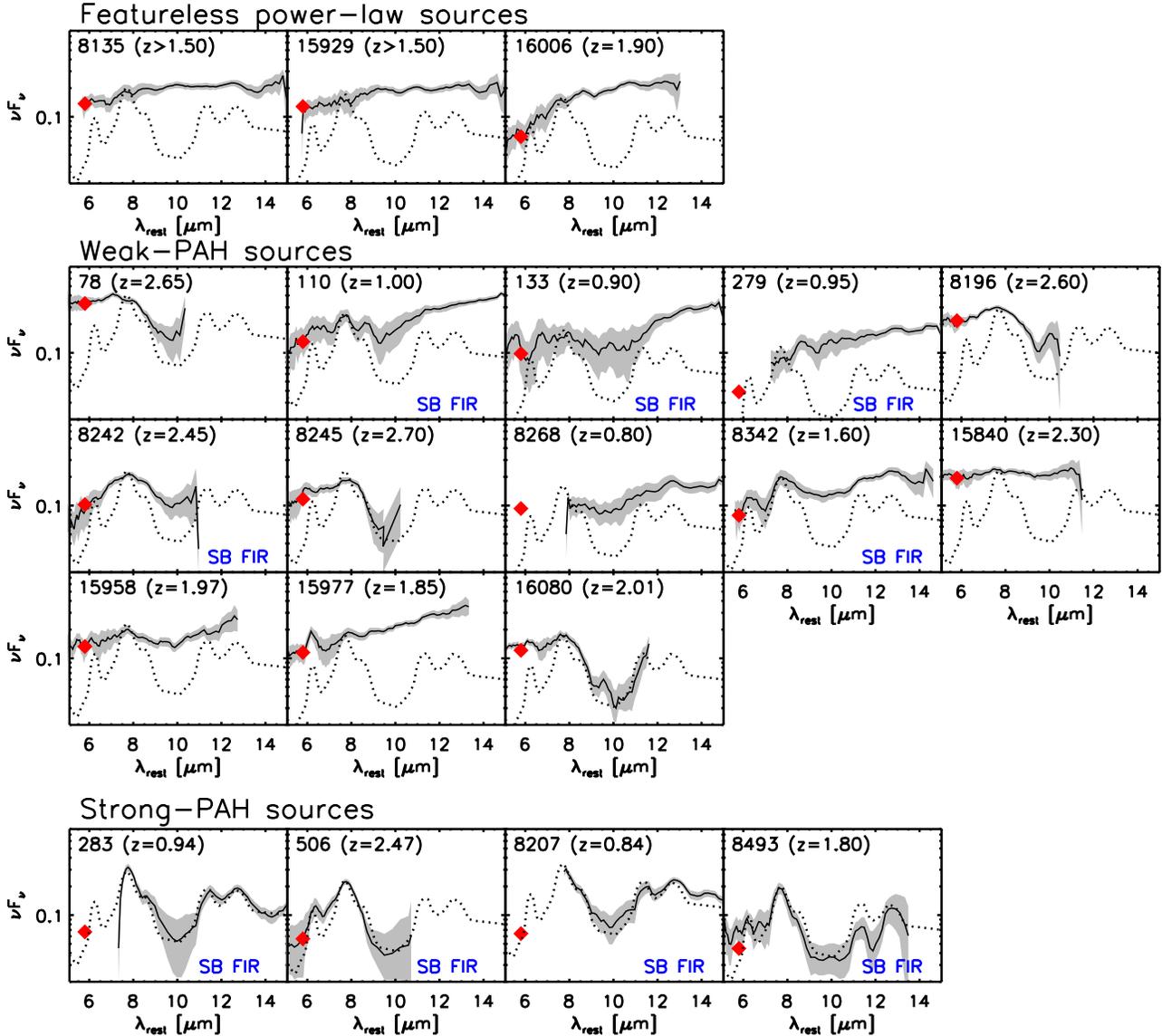}}
}
\vspace{0.05in} 
\figcaption[IRS_SEDs_color.eps]{
{\footnotesize Rest-frame {\it Spitzer} IRS spectra (solid black
curves) and $\pm1\sigma$ errors (grey regions) in units of
$10^{-15}$~W~m$^{-2}$, adapted from Fig. 3
of \citet[][]{Sajina2007}. The red diamond shows the 5.8$\mu$m
continuum flux extracted from spectral deconvolution, which for
$z\la1.5$ objects had to be extrapolated based on additional IRAC
photometry (not shown). Also shown is the local starburst galaxy
template from \citet[][]{Brandl2006}, normalized to the rest-frame IRS
data of the xFLS ULIRGs to demonstrate the maximum contribution a
typical starburst could likely make to the spectra.  Note that the
templates shown do not constitute the actual starburst contributions
derived from mid-IR or full SED deconvolutions, which are estimated to
be substantially lower in nearly all cases \citep[][and further work
in preparation]{Sajina2007, Sajina2008}. Importantly, while the
strong-PAH sources show obvious PAH features comparable to the
starburst template, the weak-PAH and featureless power-law sources are
clearly continuum dominated and look {\it nothing} like PAH-dominated
starbursts. In nearly all the latter ULIRGs, there is strong, excess
continuum at 5.8$\mu$m presumably powered by obscured AGN. As noted in
blue for individual objects, the case for AGN dominance is not as
strong in the far-IR, where nearly half (9/20) of the sample are
starburst-dominated.
\label{fig:irs_spec}}}
\vspace{0.2cm} 
\end{figure*} 

The effectiveness of the color selection is confirmed by low
resolution, mid-IR spectra \hbox{(14--40$\mu$m)} taken with {\sl
Spitzer} InfraRed Spectrograph (IRS) and optical/near-IR spectra taken
with Keck. A detailed description of the mid-IR spectra and analyses
combining spectra with far-infrared and sub-millimeter photometry have
been published in \citet[][hereafter Y07]{Yan2007}, S07, and S08. Here
we summarize the salient results for this sample.  

The mid-IR low-resolution spectra provide redshift measurements for 47
of the 52 sources: the majority (35/47\,=\,74\%) lie at
1.5\,$<$\,$z$\,$<$3\,.2, while a small fraction (12/47\,=\,26\%) lie
at 0.65\,$<$\,$z$\,$<$\,1.5. One additional source, MIPS\,279, has a
Keck redshift of $z=0.95$. The remaining four sources are all
optically faint ($R\ga25$) and have smooth power-law mid-IR spectra:
near-IR photometry allows us to estimate $z_{ph}=1.9\pm0.5$ for
MIPS\,15929 based on a spectral energy distribution (SED) template
fit, while the others are only detected in \hbox{1--2} bands and thus
likely lie at $z_{ph}\ga1.5$ \citep[e.g.,][]{Alexander2001,
Rigby2005}.

\begin{deluxetable*}{rcccccccccccccc}
\tabletypesize{\scriptsize}
\tablewidth{0pt}
\tablecaption{xFLS ULIRGs\label{tab:sample}}
\tablehead{
\colhead{ID} & 
\multicolumn{3}{c}{Net Counts} & 
\multicolumn{3}{c}{Observed F$_{\rm X}$} &
\colhead{z} & 
\colhead{Rest-frame Flux} & 
\multicolumn{3}{c}{Rest-frame Luminosity} &
\colhead{PAH} &
\colhead{Optical} &
\colhead{$\tau_{\rm 9.7\mu m}$} 
\\
\colhead{} & 
\colhead{SB} & 
\colhead{HB} & 
\colhead{FB} & 
\colhead{SB} & 
\colhead{HB} & 
\colhead{FB} & 
\colhead{} & 
\colhead{\hbox{2--10~keV}} & 
\colhead{\hbox{2--10~keV}} & 
\colhead{$5.8\,\mu{\rm m}$} &
\colhead{\hbox{3--1000$\,\mu{\rm m}$}} &
\colhead{} &
\colhead{AGN} &
\colhead{}} 
\tableheadfrac{0.05}
\startdata
\multicolumn{15}{c}{Marginal \hbox{X-ray} Detections} \\
\hline
8268  &  0.6 &  5.8\rlap{$^{w}$} &  6.4\rlap{$^{w}$} & $<$1.60 & 6.82 & 3.76             &  0.80     & $<$7.24 & $<$43.34 & 44.40 & 45.0 & w & --- & 0.8$\pm$0.4 \\ 
8342  &  0.9 &  3.6 &  4.5\rlap{$^{w}$} & $<$1.37 & $<$4.77 &  1.75 &  1.56             & $<$1.71 & $<$43.44 & 45.08 & 46.0 & w & yes & 0.2$\pm$0.4 \\ 
\hline
\multicolumn{15}{c}{\hbox{X-ray} Upper Limits} \\
\hline
78    &  1.2 &  3.6 &  4.8 & $<$1.76 & $<$3.58 & $<$5.25 &  2.65                        & $<$2.20 & $<$44.10 & 46.24 & 46.7 & w & --- & 3.4$\pm$0.5 \\ 
110   & \llap{$-$}0.4 &  3.5 &  3.2 & $<$1.06 & $<$5.29 & $<$5.34 &  1.05                        & $<$2.58 & $<$43.18 & 44.79 & 45.6 & w & yes & 0.5$\pm$0.5 \\ 
133   & \llap{$-$}1.2 & \llap{$-$}4.2 & \llap{$-$}5.5 & $<$3.12 & $<$18.30 & $<$5.49 &  0.90                        & $<$4.52 & $<$43.26 & 44.55 & 45.3 & w & yes & 0.9$\pm$ 0.8 \\ 
279   & \llap{$-$}0.1 &  0.8 &  0.8 & $<$0.89 & $<$4.40 & $<$2.43 &  0.95\rlap{$^{*}$}           & $<$1.59 & $<$42.90 & 44.50 & 45.2 & w & no & 0.0$\pm$0.1 \\ 
283   &  1.7 &  4.1 &  5.8 & $<$1.89 & $<$9.49 & $<$4.57 &  0.94                        & $<$3.13 & $<$43.15 & 44.43 & 44.9 & s & no & 1.5$\pm$0.9 \\ 
506   &  0.1 &  0.2 &  0.3 & $<$1.50 & $<$5.70 & $<$3.14 &  2.47                        & $<$1.88 & $<$43.96 & 45.40 & 46.3 & s & no & $>$6.7 \\ 
8135  &  1.0 & \llap{$-$}0.1 &  0.9 & $<$2.10 & $<$4.90 & $<$2.70 &  \llap{$>$}1.50\rlap{$^{p}$} & $<$3.76 & $<$43.73 & \llap{$>$}45.25 & \llap{$>$}46.0 & w & --- & --- \\ 
8196  &  0.0 & \llap{$-$}0.2 & \llap{$-$}0.2 & $<$0.86 & $<$2.85 & $<$1.57 &  2.59                        & $<$1.08 & $<$43.77 & 46.04 & 46.5 & w & yes & 1.3$\pm$0.4 \\ 
8207  & \llap{$-$}1.7 &  0.2 & \llap{$-$}1.5 & $<$1.13 & $<$13.13 & $<$4.12 &  0.83                        & $<$2.44 & $<$42.92 & 44.29 & 45.0 & s & no & 0.9$\pm$0.5 \\ 
8242  &  1.5 &  1.5 &  3.0 & $<$1.86 & $<$7.14 & $<$3.93 &  2.45                        & $<$3.15 & $<$44.17 & 45.65 & 46.3 & w & --- & 0.9$\pm$0.5 \\ 
8245  &  0.0 & \llap{$-$}0.1 & \llap{$-$}0.2 & $<$0.88 & $<$5.85 & $<$1.61 &  2.70                        & $<$1.11 & $<$43.82 & 45.81 & 46.4 & w & --- & 2.8$\pm$0.7 \\ 
8493  & \llap{$-$}0.7 & \llap{$-$}0.4 & \llap{$-$}1.1 & $<$1.16 & $<$4.80 & $<$3.17 &  1.80                        & $<$3.24 & $<$43.86 & 44.96 & 45.4 & s & --- & $>$4.7 \\ 
15840 &  2.6 &  0.2 &  2.9 & $<$2.14 & $<$3.42 & $<$2.35 &  2.30                        & $<$3.18 & $<$44.11 & 45.86 & 46.3 & w & --- & 0.2$\pm$0.2 \\ 
15929 &  0.0 &  0.9 &  0.8 & $<$0.97 & $<$4.80 & $<$2.64 &  \llap{$>$}1.50\rlap{$^{p}$} & $<$2.46 & $<$43.54 & \llap{$>$}45.22 & \llap{$>$}45.9 & w & --- & --- \\ 
15958 &  0.9 &  0.9 &  1.7 & $<$1.35 & $<$4.70 & $<$2.59 &  1.97                        & $<$3.07 & $<$43.93 & 45.53 & 46.1 & w & --- & 0.5$\pm$0.4 \\ 
15977 &  3.9 & \llap{$-$}3.4 &  0.5 & $<$2.36 & $<$2.53 & $<$3.48 &  1.85                        & $<$3.33 & $<$43.90 & 45.40 & 46.2 & w & --- & 0.1$\pm$0.1 \\ 
16006 & \llap{$-$}0.8 &  0.9 &  0.1 & $<$1.14 & $<$9.43 & $<$4.16 &  1.90\rlap{$^{p}$}           & $<$2.04 & $<$43.72 & 45.16 & 45.9 & w & --- & --- \\ 
16080 & \llap{$-$}0.2 & \llap{$-$}0.1 & \llap{$-$}0.3 & $<$0.89 & $<$5.91 & $<$1.63 &  2.01                        & $<$1.60 & $<$43.69 & 45.51 & 46.2 & w & yes\rlap{$^{\dagger}$} & 2.1$\pm$0.5 \\ 
\hline
\multicolumn{15}{c}{\hbox{\hbox{X-ray}} Stacking Results} \\
\hline
0.8$\le$$z$$<$1.5 & -2.1 &  4.5 &  2.4 & $<$0.26 & $<$3.15 & $<$1.53 &  $\langle$0.96$\rangle$ & $<$0.65 & $<$42.44 & $\langle$44.53$\rangle$ & $\langle$45.3$\rangle$ & --- & --- & --- \\
1.5$\le$$z$$<$3.0 &  4.3 &  3.6 &  8.0 & $<$0.22 & $<$0.67 & $<$0.46 &  $\langle$2.09$\rangle$ & $<$0.28 & $<$42.95 & $\langle$45.49$\rangle$ & $\langle$46.1$\rangle$ & --- & --- & --- 
\enddata
\tablecomments{
{\it Col. (1)} MIPS source number, from \citet{Yan2007}. Last two entries denote stacking results (see
$\S$\ref{data}).
{\it Cols. (2--4)} Total background-subtracted counts in the
\hbox{0.5--2~keV} (SB), \hbox{2--8~keV} (HB), and \hbox{0.5--8.0~keV} (FB) bands,
respectively, as measured by {\sc acis extract} in 90\% encircled
energy region at 1.49~keV. 'w' denotes a source detection with {\sc
wavdetect} at the $10^{-5}$ significance threshold.
{\it Cols. (5--7)} Flux in the \hbox{0.5--2~keV}, \hbox{2--8~keV}, and
\hbox{0.5--8.0~keV} bands, respectively, in units of
10$^{-15}$~erg~s$^{-1}$~cm$^{-2}$ assuming a power-law model with
$\Gamma=1.4$. If the source is undetected, we instead provide the
3$\sigma$ upper limit using the Bayesian determination
of \citet{Kraft1991}.
{\it Col. (8)} Redshift, from S08. ``$p$'' indicates a
photometric redshift. Note that MIPS 279 (``*'') was quoted in early
xFLS papers as having a redshift of 1.23; this as since been
revised to the value adopted in the table.
{\it Col (9)} Observed rest-frame \hbox{2--10~keV} flux in units of
10$^{-15}$~erg~s$^{-1}$~cm$^{-2}$ assuming a power-law model with
$\Gamma=1.4$.
{\it Col (10)} Logarithm of the observed rest-frame \hbox{2--10~keV}
luminosity in units of erg~s$^{-1}$.
{\it Col. (11)} Logarithm of the rest-frame 5.8\,$\mu$m continuum
luminosity, from S07, in units of erg~s$^{-1}$.
{\it Col (12)} Logarithm of the rest-frame \hbox{3--1000\,$\mu$m} continuum
luminosity, from S08, in units of erg~s$^{-1}$.
{\it Col (13)} Relative strength of PAH features in IRS spectrum,
from S07. Strong (``s'') PAH sources have
EW(7.7$\mu$m)$>$0.8$\mu$m, while weak (``w'') PAH sources have lower EWs.
{\it Col. (14)} Presence of AGN as assessed by optical spectroscopy,
from S08. MIPS\,16080 (``$^{\dagger}$'') has starburst-like line
ratios, but an asymmetric [O\,{\sc III}] profile typical of an AGN
outflow; this is similar to brighter, radio-excess
ULIRGs \citep[e.g.,][]{Buchanan2006}.
{\it Col. (15)} Silicate absorption at 9.7$\mu$m in dimensionless units.}
\end{deluxetable*}

At such redshifts, most of the mid-IR spectra cover the rest-frame
\hbox{$\sim$\,5\,--\,15\,$\mu$m} band, with a minimum signal-to-noise per
pixel of 4; for convenience, Fig.~\ref{fig:irs_spec} reproduces these
spectra from \citet{Sajina2007} for all of the xFLS ULIRGs listed in
Table~\ref{tab:sample}. Locally, the PAH strength has been shown to be
a robust tracer of star formation \citep[e.g.,][]{Brandl2006,
ODowd2009} and low-equivalent width sources are virtually all
AGN-dominated \citep[at least in the mid-IR; e.g.,][]{Genzel1998,
Lutz1998, Tran2001, Farrah2007, Desai2007}. Despite the modest
signal-to-noise for some spectra, comparison to maximal contributions
from an average starburst template of \citet{Brandl2006} in
Fig.~\ref{fig:irs_spec} clearly demonstrates that the vast majority of
xFLS ULIRGs studied here are continuum-dominated. We highlight that
aside from the four starburst-dominated objects, the best-fitted
starburst contributions to the mid-IR are generally much lower than
the maximal contribution shown.

When the entire xFLS sample are more rigorously decomposed into PAH,
continuum, and obscuration spectral components, S07 find that
$\sim$\,75\% of the sample are weak-PAH sources with
\hbox{$EW_{rest}({\rm 7.7\,\mu m\,PAH}) \la 0.8\,\mu$m}, consistent with
AGN-dominance in the mid-IR. Although the remaining 25\% appear to be
starburst-dominated, the S07 template deconvolution indicates that
only $\sim$50\% of the 5.8$\mu$m continuum can reasonably be
attributed to star formation, prompting the need for an additional
hot-dust component from a dust-obscured AGN. These results are
reinforced by the known correlation between starburst PAH strength and
the host galaxy stellar bumps \citep[e.g.,][]{Lacy2004, Stern2005,
Weedman2006, Teplitz2007}, wherein strong-PAH sources generally show
clear stellar bumps in the IRAC bands while weak-PAH sources show no
such bumps and usually are just power laws (S07). Subsequent analysis
of the optical spectra, radio properties, and UV-to-radio SEDs by S08
further confirm the mid-IR spectral deconvolutions, concluding that
strong-PAH sources typically have AGN contributions
of \hbox{$\sim$20--30\%} of the total $L_{\rm IR}$, while weak-PAH
sources generally have AGN contributions of $\ga$70\%. Notably,
weak-PAH sources are roughly twice as likely to lie at $z\ga1.5$ and
have a substantially higher fraction of optically-identified AGN
compared to strong-PAH sources (see Fig.~\ref{fig:irs_spec} and
S08). Thus the conclusions regarding $z$$\sim$2 ULIRGs should be
robust.

We additionally note that Sajina et al. (2010, in prep.) study the
UV-to-radio SEDs of a large sample of 24um-selected sources including
most of the objects discussed here. This work includes comparison with
local galaxy templates as well as fits to composite
quasar$+$starburst templates.  They find that such fits are not
sensitive to the details of the IRS spectra but rather to the overall
1--1000$\mu$m SED shapes of these sources. They conclude that although
weak-PAH sources can still have their total IR luminosities dominated
by starburst activity, the 5.8$\mu$m continuum of such sources is
overwhelmingly dominated by obscured quasars (e.g.,
Fig.~\ref{fig:irs_spec}). This conclusion supports those
of \citet{Polletta2008}, whose sample also includes a few of the
sources studied here, but where the fitting includes theoretical torus
models rather than a quasar template. Both studies suggest that unless
$z$$\sim$2 starbursts and AGN have dramatically different SED shapes
than has been hitherto observed, the conclusion that weak-PAH sources
have 5.8$\mu$m continua dominated by AGN is sound.

The total IR luminosities of the sample, derived from the mid-IR
spectrum plus 70$\mu$m, 160$\mu$m, and 1.2mm photometry (S08), range
from \hbox{(0.04--2)$\times10^{13}$~$L_{\odot}$}, with a median of
\hbox{5$\times10^{12}$~$L_{\odot}$}. While of comparable bolometric 
luminosities to QSOs, these sources are typically much more obscured,
as evidenced by their steeper mid-IR slopes and often significant
9.7\,$\mu$m silicate absorption (e.g., mean optical depth
$\langle\tau_{9.7}\rangle$\,$\sim$\,1.4). This latter quantity,
$\tau_{9.7}$, appears to correlate strongly with X-ray derived $N_{\rm
H}$ \citep[][]{Shi2006}. We note that the distribution of slope
and absorption values are somewhat more extreme than those typical of
type II AGN \citep[e.g.,][]{Desai2007, Hao2007}, suggesting that these
objects may not merely be QSOs observed at unfavorable viewing
angles \citep[e.g., the AGN unification model;][]{Antonucci1993}, but
might represent a particular obscured class or evolutionary phase of
QSOs.

Clearly there could be variations in the individual starburst and AGN
mid-IR SEDs, and thus the accuracy of the spectral decompositions is
somewhat limited. However, this is unlikely to change the majority of
5.8$\mu$m continuum measurements by more than 10--20\% typically, and
thus would have little effect on our results. We also caution that the
quoted rest-frame mid-IR luminosities from S07 have not yet been
corrected for intrinsic mid-IR absorption, which is difficult to
determine on a source-by-source basis. In particular, some fraction of
our sources have red continua indicative of strong mid-IR extinction
but no obvious accompanying silicate absorption, and thus may have
considerable variation in either their dust geometry or intrinsic
extinction law. Although our quoted continuum at 5.8\,$\mu$m lies near
a minimum in the mid-IR extinction curve \citep[e.g.,][]{Chiar2006},
the impact of dust could still be considerable. With extinction
typical of the Galactic Center (GC), for instance, our $\tau_{9.7}$
measurements could equate to a 5.8\,$\mu$m flux decrement of $\sim$1.6
on average and $\sim$10 in a few extreme cases. Additionally, the
presence of molecular absorption features due to water ice and
hydrocarbons often accompany strong silicate absorption and may result
in additional intrinsic absorption (e.g., around 6.2$\mu$m) in some
weak-PAH sources \citep[e.g.,][]{Spoon2002, Spoon2004, Imanishi2006b,
Imanishi2008, Risaliti2006, Sani2008}. Care has been taken to exclude
affected spectral regions but such features may still impact the
decomposition fitting. \citet{Sajina2009} explores in detail some
instances of water ice and hydrocarbons within the xFLS sample, to
which we refer interested readers. Both features could lead to
significant underestimatates of the AGN continuum, which must thus be
regarded as lower limits to their intrinsic values. Finally, objects with
$z\la1.5$ have somewhat larger uncertainties in $L_{\rm 5.8\mu m}$
because estimates can only be extrapolated based on IRS spectral
decomposition above rest-frame \hbox{$\sim$6--8$\mu$m} and broadband
IRAC photometry below.

Thus to summarize the strong evidence for dominant AGN activity in
the xFLS sample:

\begin{itemize}

\item Based on spectral deconvolution of the IRS spectra, 
$\approx$75\% of the sample are continuum-dominated, weak-PAH objects
powered by AGN in the mid-IR, and even among strong-PAH objects,
additional AGN continuum components are generally required. While
there clearly could be some uncertainty in the deconvolution, the
possible issues (e.g., templates, extinction from dust and ice)
largely skew toward underestimating the true $L_{\rm 5.8\mu m}$
continuum of the xFLS ULIRGs.

\item Spectral decomposition of the rest-frame UV-to-radio SEDs confirm
the modeling results of the mid-IR data alone, suggesting that
strong-PAH sources typically have $\sim$20--30\% AGN fractions of
$L_{\rm IR}$, while weak-PAH sources by contrast tend to have
$\ga$70\% AGN fractions, with a few outliers having comparable
contributions from AGN and starbursts.

\item Radio and, where available, optical-line diagnostics support the 
presence of AGN in $\sim$60\% of the $z$$>$1.5, predominantly weak-PAH
ULIRGs, independent of any IR-based diagnostics.

\end{itemize}

To explore the \hbox{X-ray} energetics and constrain the nature of any
potential \hbox{X-ray} absorption in the above sources, we
obtained \hbox{X-ray} observations for a random subset of the full
sample.

\section{Observational Data and Reduction Methods}\label{data}

We observed 20 of the 52 xFLS ULIRGs with five {\it Chandra} ACIS-I
pointings of 30~ks each (PI: Yan; Obsids: 
\dataset [ADS/Sa.CXO#obs/07824] {7824}, 
\dataset [ADS/Sa.CXO#obs/07825] {7825},
\dataset [ADS/Sa.CXO#obs/07826] {7826},
\dataset [ADS/Sa.CXO#obs/07827] {7827}, and
\dataset [ADS/Sa.CXO#obs/07828] {7828}). 
The data were processed following standard procedures using {\sc ciao}
(v3.4) software.\footnote{http://asc.harvard.edu.} We additionally
removed the 0\farcs5 pixel randomization, corrected for charge
transfer inefficiency, performed standard {\it ASCA} grade selection,
excluded bad pixels and columns, and screened for intervals of
anomalous background using a threshold of 3$\sigma$ above or below the
mean (this excluded at most \hbox{1--2~ks} for each observation). Further
analysis was performed on reprocessed {\it Chandra} data, using {\sc
ciao}, {\sc ftools} (v6.3) and custom software including {\sc acis
extract} (v3.172)\footnote{
http://www.astro.psu.edu/xray/docs/TARA/ae\_users\_guide.html}

Source detections were determined using the {\sc wavdetect}
source-searching algorithm with a $10^{-5}$ significance
threshold. While this threshold is typically considered 'lenient'
(i.e., \hbox{10--20} false sources over full {\it Chandra} ACIS-I
field), the low xFLS ULIRG source density means we are only interested
in detections at a handful of locations over the ACIS-I field-of-view,
and thus it translates here into a very robust search
significance. Since our sources are likely to be obscured, we searched
in the standard full \hbox{(0.5--8.0~kev)},
soft \hbox{(0.5--2.0~keV)}, and hard \hbox{(2--8~keV)} bands
separately. We additionally searched in the observed \hbox{1--4~keV}
band, which roughly equates to rest-frame \hbox{2--10~keV} at $z>1.5$,
to determine the assignment of detections and upper limits at the
rest-frame
\hbox{2--10~keV} fluxes assessed below. There are typically
\hbox{$\sim$90--100} \hbox{X-ray} sources detected per observation, which we
matched to optical counterparts from the xFLS optical survey
catalog \citep{Fadda2004}, providing $\sim$30 highly solid matches
within 0\farcs5 per observation. The matches allowed registration of
the \hbox{X-ray} astrometric frame to the optical, with typical linear
shifts of \hbox{0\farcs1--0\farcs2} to the original \hbox{X-ray} frame. The
resulting 1$\sigma$ registration residuals are $\approx$0\farcs3,
ensuring spatially robust \hbox{X-ray} identifications and upper
limits. Among the 20 objects observed, only two were formally detected
in any of the above bands (see Table~\ref{tab:sample}).

Aperture-corrected photometry was performed using \hbox{\sc acis
extract} with a 90\% encircled-energy region derived from the {\it
Chandra} PSF library, with 3$\sigma$ upper limits calculated
following \citet{Kraft1991}. Fluxes were derived assuming a power-law
model with $\Gamma=1.4$ consistent with the spectrum of the CXRB. We
also performed photometry on specific sub-bands to allow more accurate
constraints on rest-frame fluxes at the redshifts of our sources. To
this end, we use the observed \hbox{0.5--2.0~keV} and \hbox{1--4~keV}
bands to estimate the rest-frame \hbox{2--10~keV} fluxes (or upper
limits) and subsequent luminosities for objects with $z\la1.5$ and
$z>1.5$, respectively. An additional K-correction of
order \hbox{15--20\%} is applied to account for individual redshifts.

Regarding the two detected sources, we find that one is detected in
the full and hard bands, while the other only in the full
band. Although the photon statistics here are poor \citep[both sources
are formally consistent with zero net counts at \hbox{1--2$\sigma$}
confidence following][]{Gehrels1986}, the photon energy distributions
lean toward both being heavily obscured. An unabsorbed AGN with a
$\Gamma=1.9$ power-law spectrum, for instance, would have three times
as many \hbox{0.5--2.0~keV} counts as \hbox{2--8~keV} counts, easily
detectable in our {\it Chandra} observations. Given the effective area
of {\it Chandra}, a large neutral hydrogen column of
\hbox{$\sim$$10^{23}$--$10^{24}$~cm$^{-2}$} would be required to arrive at
the observed count distributions of both sources. Neither source is
formally detected in the rest-frame \hbox{2--10~keV} bands described
above, however, so we quote their 3$\sigma$ upper limits for this
flux. Assuming $\Gamma=1.4$ again ($\Gamma=1.9$ would change these
values by only $\approx$10\%), the observed full-band luminosities of
MIPS\,8268 and MIPS\,8342 are \hbox{$1.0\times10^{43}$~erg~s$^{-1}$} and
\hbox{$3.2\times10^{43}$~erg~s$^{-1}$}, respectively, and extrapolate to
rest-frame \hbox{2--10~keV} luminosities of
\hbox{$7.8\times10^{42}$~erg~s$^{-1}$} and \hbox{$2.5\times10^{43}$~erg~s$^{-1}$}
for the two objects.

To place stronger average constraints on our sample, we also
performed \hbox{X-ray} stacking analyses on the undetected sources,
which has been successfully employed on numerous source populations in
both wide and deep field X-ray surveys \citep[e.g.,][]{Brandt2001,
Brand2005, Daddi2007b, Lehmer2008}. We divided the 20 ULIRGs into two
subsets based on redshift (0.8$\la$$z$$<$1.5 and 1.5$\la$z$<$3.0),
taking care to exclude both the two detected sources and the three
sources which lie on ACIS chip S7, since the extended PSFs from the
latter include too much background to improve the signal-to-noise. For
consistent aperture corrections, we again used the 90\%
encircled-energy aperture region measured at 1.49~keV from {\sc acis
extract}. For rest-frame luminosity estimates, we stacked counts only
from the \hbox{2--10~keV} rest-frame bandpasses of the sources (i.e.,
in the observed \hbox{0.5--2.0~keV} and \hbox{1.0--4.0~keV} bands for
z>1.5 and z<1.5, respectively). Aside from the use of individual
geometric source apertures instead of circular apertures, our method
largely follows that of \citet[][and references
therein]{Lehmer2008}. To properly account for spatial variations in
pixel sensitivity due to chip gaps, bad pixels, and vignetting, the
total number of background counts within our stacked aperture was
determined by scaling the cumulative background counts found within
local annuli by the ratio of the summed exposure times in the source
and background regions, respectively. No significant detection was
found in either redshift subsample,\footnote{The most significant
detection was only $\approx$1.7$\sigma$ in the full-band for the
1.5$\la$z$<$3.0 subsample.} so we calculated 3$\sigma$ upper limits
following \citet{Kraft1991}. These count limits were converted to
\hbox{2--10~keV} rest-frame fluxes using the previously adopted spectral
model for photometry above.

\section{Discussion}\label{discussion}

We would like to use the above constraints to understand the nature of
the xFLS ULIRGs and their context with respect to the rest of the
high-z AGN population. Continuum emission at hard \hbox{X-ray} and
mid-infrared wavelengths, as well as emission from the narrow-line
region, are widely considered to provide the most robust constraints
on AGN bolometric luminosities \citep[e.g.,][]{Bassani1999, Xu1999,
Lutz2004, Heckman2005, Imanishi2006a, Melendez2008, Nardini2008,
Vega2008, McKernan2009}, as each offers a unique measure of the
primary AGN energy output over large dynamic
range \hbox{($\sim$4--5~dex)} before contamination from host star
formation sets in. The dependence of AGN properties on orientation and
intrinsic obscuration is a long-standing problem when trying to assess
the true power of AGNs \citep[e.g.,][]{Antonucci1993} and none of
these tracers are without its faults. However, because these tracers
are affected differently by orientation and obscuration, we can assess
the nature of obscuration present in our xFLS ULIRGs by comparing
their \hbox{X-ray} upper limits to other less obscured tracers such as
rest-frame 5.8$\mu$m continuum and (in a few cases) [O\,{\sc III}]
flux. This is essentially the same approach adopted
by \citet[][hereafter A08]{Alexander2008} to constrain Compton-thick
AGN in the GOODS-N region. First, however, we would like to assess the
applicability of these relationships to ULIRGs using several bright,
well-characterized objects from the literature.

\begin{figure*}[t]
\vspace{0.1in}
\centerline{
\hglue-0.2cm{\includegraphics[width=18.8cm]{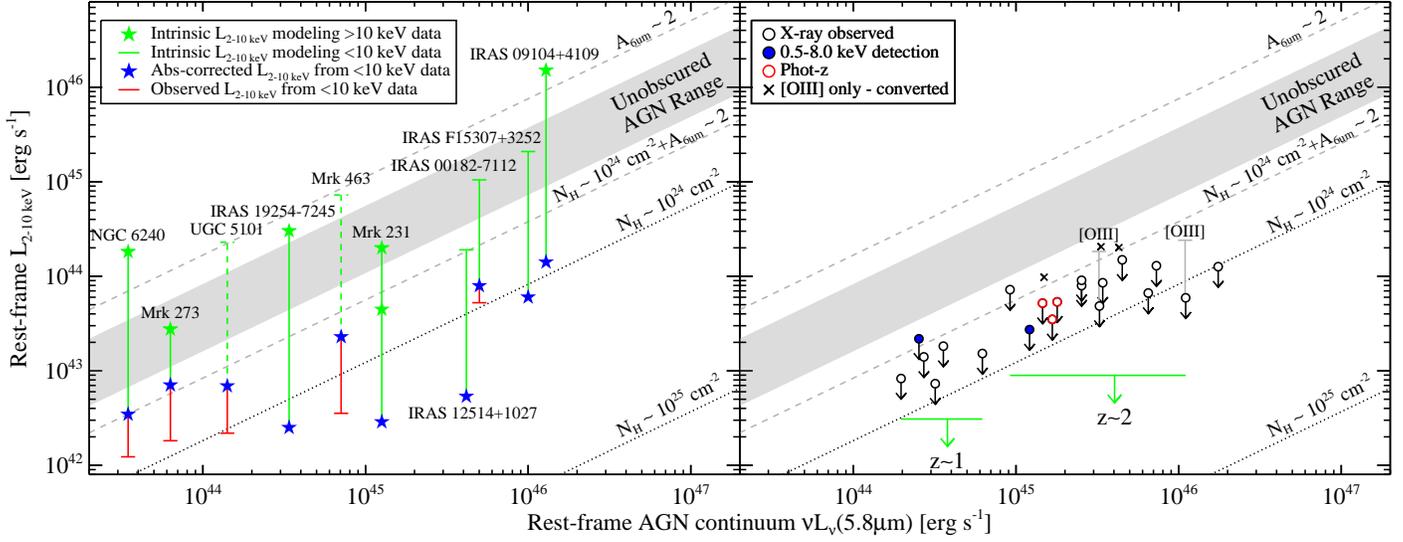}}
}
\vspace{-0.0in} 
\figcaption[LIR_vs_Lx.eps]{
{\footnotesize Rest-frame \hbox{2--10~keV} luminosity versus 5.8\,$\mu$m AGN
continuum luminosity for several bright ULIRGs from published
literature thought to host Compton-thick or nearly Compton-thick AGN
({\it left}) and the {\it Chandra}-observed xFLS ULIRGs ({\it right}).
All 5.8\,$\mu$m AGN luminosities were calculated from the best-fitting
AGN component via spectral deconvolution (e.g., S07) and have not been
corrected for mid-IR absorption. In both panels, the grey shaded
region denotes the 1$\sigma$ scatter in
rest-frame \hbox{2--10~keV-to-5.8\,$\mu$m} luminosity ratio range
found for type~1 AGNs (Bauer et al., in preparation), the black dotted
lines demonstrate the effects of \hbox{X-ray} absorption by column
densities of \hbox{$N_{\rm H}\sim10^{24}$~cm$^{-2}$} \citep[using the
model presented in][]{Alexander2005b} and \hbox{$N_{\rm
H}\sim10^{25}$~cm$^{-2}$} \citep[assuming the \hbox{X-ray} spectrum of
NGC 1068;][]{Bassani1999}, and the grey dashed lines show the effect
of mid-IR extinction by \hbox{$A_{\rm 6\mu m}\sim2$~mag} to unobscured
and \hbox{$N_{\rm H}\sim10^{24}$~cm$^{-2}$} sources \citep{Chiar2006}.
For the bright, individually-labeled ULIRGs ({\it left}): red lower
limits denote absorbed, rest-frame $L_{\rm 2-10\,keV}$ values (i.e,
the observed data; note the composite spectra for some objects were
too contaminated by star formation at low energies to assess the AGN
contribution or apparent absorption unambiguously); blue stars denote
absorption-corrected, rest-frame $L_{\rm 2-10\,keV}$ values derived
from the best-fitted models to $<10$~keV spectra, assuming the data
are predominantly direct continuum; green stars denote
absorption-corrected, rest-frame $L_{\rm 2-10\,keV}$ values derived
from the best-fitted models to both $<10$~keV and $>$10~keV spectra,
assuming the $>$10~keV data is largely direct continuum while the
$<10$~keV data is only reflected/scattered continuum; and green upper
limits are absorption-corrected, rest-frame $L_{\rm 2-10\,keV}$
extrapolations derived from the best-fitted models to $<10$~keV
spectra, assuming the data is reflected/scattered continuum only. Note
that all luminosities were taken from the literature.
For the xFLS ULIRGs ({\it right}): \hbox{X-ray} upper limits (3$\sigma$) are
calculated following \citet{Kraft1991}, assuming $\Gamma=1.4$; blue
points indicate {\sc wavdetect} full-band detections; red points
denote sources with photometric redshift constraints; and green bars
represent the average stacked upper limits for our $z\sim1$ and
$z\sim2$ subsamples. The individual upper limits lie close to the
\hbox{$N_{\rm H}\sim10^{24}$~cm$^{-2}$} track, demonstrating that most xFLS
ULIRGs are highly obscured, while the stacked averages imply that a
significant fraction are likely to be at least mildly
Compton-thick. For the five xFLS ULIRGs with observed [O\,{\sc III}]
constraints, we have converted the [O\,{\sc III}] luminosities
(uncorrected for extinction) to equivalent absorption-corrected \hbox{X-ray}
luminosities using the correlation from Bauer et al. (2009, in
preparation); those observed by {\it Chandra} are plotted as grey
upper bars, while the rest are shown as crosses.
\label{fig:lir_lx}}}
\vspace{0.2cm} 
\end{figure*}

\subsection{Guidance from IR-Bright ULIRGs}\label{brightnature}

A correlation between \hbox{2--10~keV} and 5.8$\mu$m luminosity has
now been well-established for both type 1 and absorption-corrected
type 2 AGN \citep[e.g.,][]{Lutz2004, Sturm2006}. While it is not yet
clear why this correlation is as tight as observed \citep[e.g., the
apparent absorption-corrected \hbox{2--10~keV} luminosities of type 2
AGN can often be substantially lower than extrapolations based on
$>$10~keV luminosities;][]{Heckman2005, Melendez2008}, or what
functional form best describes the correlation
is \citep[e.g.,][]{Lutz2004, Fiore2008, Lanzuisi2009}, empirically
this correlation implies that AGN are relatively robust, immutable,
and scalable physical systems. In both panels of Fig.~\ref{fig:lir_lx}
we show the intrinsic rest-frame \hbox{2--10~keV-to-5.8\,$\mu$m} range
for type~1 AGNs (i.e., only continuum-dominated mid-IR sources with
little or no PAH emission), where the correlation is given as
\hbox{$L_{\rm 5.8\mu m}=10^{-8.7\pm2.6}L_{\rm 2-10\,keV}^{1.21\pm0.06}$},
with both luminosities in units of erg~s$^{-1}$, and the grey shaded
region denotes 1$\sigma$ luminosity range (Bauer et al., in
preparation); we note that this fit is consistent with the original
correlation found by \citet{Lutz2004} at low luminosities and lies
intermediate between the extrapolations of \citet{Lutz2004}
and \citet{Lanzuisi2009} at high mid-IR luminosities.

To demonstrate how \hbox{X-ray} absorption and mid-IR extinction affect the
intrinsic AGN output, we also show in Fig.~\ref{fig:lir_lx} the
expected ratios for a typical AGN (taken as the center of the
unobscured AGN range) assuming (1) the 5.8$\mu$m continuum is
intrinsic but the \hbox{X-ray} emission is absorbed by a column densities of
\hbox{$N_{\rm H}\sim10^{24}$~cm$^{-2}$} and \hbox{$N_{\rm
H}\sim10^{25}$~cm$^{-2}$},\footnote{The value of \hbox{$N_{\rm
H}\sim10^{24}$~cm$^{-2}$} is roughly where obscuration is expected to
start becoming Compton-thick, and has been calculated using a model
similar to that presented in \citet{Alexander2005b}
and \citet{Gilli2007}, wherein the primary radiation is obscured and
we only detect reflection and scattering components that typically
comprise $\approx$5\% of the intrinsic \hbox{2--10~keV}
emission. By \hbox{$N_{\rm H}\sim10^{25}$~cm$^{-2}$}, the source is
expected to be 'fully' Compton-thick such that no primary continuum
escapes, and has been calculated assuming the best-fitted \hbox{X-ray}
spectrum of NGC\,1068 from \citet{Bassani1999}.} and (2)
the \hbox{X-ray} emission is intrinsic but the 5.8$\mu$m continuum is
extincted by $A_{\rm 6\mu m}\sim2$~mag,\footnote{If we combine
the \citealp{Cardelli1989} optical/near-IR and \citealp{Chiar2006}
mid-IR extinction curves, $A_{\rm 6\mu m}\sim2$~mag is equivalent to
$A_{\rm V}\sim36$~mag assuming $R_{\rm V}=5$ appropriate for dense
starbursts/H\,{\sc II} regions or $A_{\rm V}\sim42$~mag assuming
$R_{\rm V}=3.1$ appropriate for Milky Way interstellar material,
respectively.} as might be the case in many ULIRGs. Both decrements
and their associated errors contribute to the observed scatter of this
relation, and thus it is natural to wonder how ULIRGs, which are some
of the most dusty and obscured objects in the Universe, fit into the
above picture.

To this end, we assembled a sample of IR-bright ULIRGs from the
literature, all of which have high quality {\it Spitzer} IRS
and \hbox{2--10~keV} spectra and are widely considered to host
Compton-thick AGN. This list is regrettably small because such objects
are typically \hbox{X-ray} faint and difficult to
constrain. Importantly, half of the sample have published \hbox{X-ray}
detections above 10~keV with {\it Suzaku} or {\it BeppoSAX}, allowing
the best assessment of the apparent direct \hbox{X-ray} continuum to
date. Basic properties of this sample are listed in
Table~\ref{tab:brightsample}.

The observed and intrinsic \hbox{X-ray} luminosities of these IR-bright
ULIRGs were taken directly from the references in
Table~\ref{tab:brightsample}; the \hbox{X-ray} data reduction and spectral
fitting rely on the assumptions detailed therein.  For Mrk\,231, we
list both of the intrinsic \hbox{X-ray} luminosities provided
by \citet{Braito2004}, depending on whether the \hbox{X-ray} spectrum is
dominated by scattering (low value) or reflection (high value). For
both Mrk\,463 and UGC\,5101, it has been argued that the low
observed Fe K$\alpha$ equivalent widths \hbox{($\sim$200--400~eV)}
demonstrate that the direct AGN continuum is being observed. If
scattering dominates over reflection, however, such equivalent widths
can still be fully consistent with Compton-thick
AGN \citep[e.g.,][]{Murphy2009}, as in fact appears to be the case for
Mrk\,231. Thus for these two sources, we compute intrinsic \hbox{X-ray}
luminosities assuming the observed continua are reflection/scatter
dominated and comprise $\sim$5\% of their intrinsic
values \citetext{e.g., \citealp{Alexander2005b,Gilli2007}, hereafter
G07; \citealp{Molina2009}}; these corrections are denoted with dashed
green upper limits in the left panel of Fig.~\ref{fig:lir_lx} since
they are more tentative and open to interpretation.

The 5.8$\mu$m AGN continuum luminosities for this bright ULIRG sample
were derived following the procedure used in S07. Specifically, IRS
low-resolution spectra in the form of pipeline-processed, Basic
Calibrated Data (BCDs) were downloaded from the {\it Spitzer}
archive. Sky backgrounds were generated by subtracting BCD images at
two different node positions. Flux- and wavelength-calibrated 1-D
spectra were then extracted from the sky-subtracted 2-D, BCD images
using the SSC spectral extraction software, SPICE
(v2.2).\footnote{http://ssc.spitzer.caltech.edu/postbcd/spice.html}
Finally, four segment spectra covering \hbox{7--35$\mu$m} were are fit
together to eliminate bad pixels at the edge of each order. The
resulting spectra were corrected for Galactic extinction and fit with
a power-law continuum and PAH template of M\,82.  Given that M\,82 has
one of the strongest observed mid-IR dust continua, our constraints on
AGN hot-dust components should be considered relatively
conservative. In cases where prominent 6$\mu$m ice absorption is
present, we masked it out and fit only the continua on either
side. The $L_{\rm 5.8\mu m}$ values are the monochromatic rest-frame
luminosities read off from the best-fit models. In all cases except
NGC\,6240, the continua are dominated by a power-law component. This
is also the case for our high-z xFLS sample, which is largely composed
of weak-PAH sources.  As discussed in $\S$\ref{sample}, determining
the starburst contribution to $L_{\rm 5.8\mu m}$ in the few strong-PAH
sources where it may be significant is complicated by the highly
uncertain levels of hot dust emission that can be associated with pure
starbursts. Using M\,82 as a template, for instance, we estimate that
up to half of $L_{\rm 5.8\mu m}$ in NGC\,6240 could be due to its
starburst. A starburst contribution of $\la$30\% can also be seen in
Mrk\,273 and UGC\,5101, but is negligible in the rest of the IR-bright
ULIRG sample.

The observed \hbox{2--10~keV} luminosities of the IR-Bright Compton-thick
ULIRGs in the left panel of Fig.~\ref{fig:lir_lx} all lie around the
\hbox{$N_{\rm H}\sim10^{24}$~cm$^{-2}$} track, confirming our expectation
that they are highly obscured. Once the effects of known \hbox{X-ray}
absorption are accounted for (green limits), however, these
Compton-thick ULIRGs generally lie on or above the type\,1 AGN
correlation. Importantly, the case for large \hbox{X-ray} corrections
is strongest for the sources with solid $>$10~keV \hbox{X-ray}
detections (green stars), which have the least amount of ambiguity
regarding their intrinsic \hbox{X-ray} luminosities.

Considerable uncertainties in the mid-IR also remain from the poorly
constrained gas-to-dust ratios and dust geometries in
ULIRGs. Theoretical arguments suggest that the mid-IR emission can
perhaps vary by up to an order of magnitude depending on the
distribution and composition of the obscuring
dust \citep[e.g.,][]{Pier1992, Nenkova2002, Nenkova2008}. If the
gas-to-dust ratios were substantially lower for ULIRG AGN, for
instance, then the mid-IR luminosity constraints could overestimate
the true bolometric power of the AGN. Such a correction might move
some ULIRGs into better alignment with the typical type~1 range, but
would make current discrepancies with others even more
extreme. Particular dust geometries/orientations, on the other hand,
could lead to a few magnitudes of extinction even at mid-IR
wavelengths and cause us to underestimate the true mid-IR luminosities
from these objects. The $A_{\rm 6\mu m}\sim2$~mag track demonstrates the
potential degree of such an effect. 

The left panel of Fig.~\ref{fig:lir_lx} illustrates that the known
correlation between intrinsic rest-frame 5.8$\mu$m and \hbox{2--10~keV}
emission in AGN does effectively extend to ULIRGs and that observed
\hbox{2--10~keV} decrements should provide realistic diagnostics of
obscuration in all objects containing AGN.  Settling on a sensible
criteria for selecting Compton-thick AGN, however, is not clearcut. A
conservative approach is to adopt the \hbox{$N_{\rm
H}\sim10^{24}$~cm$^{-2}$} track as our Compton-thick criteria. While
this neglects the potential mid-IR extinction that is likely to
accompany the \hbox{X-ray} absorption, it ensures that the selected
sample of candidates will have negligible contamination from less
obscured sources. Under this scheme, only six of the above 10
widely-regarded Compton-thick IR-bright ULIRGs would be selected,
demonstrating that we might miss a considerable fraction of likely
Compton-thick AGN with such an approach. On the other hand, adopting
the more inclusive \hbox{$N_{\rm H}\sim10^{24}$~cm$^{-2} + A_{\rm 6\mu
m}\sim2$~mag} track as our Compton-thick criteria would select all of
the IR-bright ULIRGs, but may potentially select AGN which are
somewhat less heavily-obscured as well. Clearly there are merits and
drawbacks to each criteria, so we will employ both in the next section
to gauge the nature of the xFLS ULIRGs.

\subsection{Constraints on the Nature of xFLS ULIRGs}\label{xflsnature}

In the right panel of Fig.~\ref{fig:lir_lx}, we compare rest-frame
5.8\,$\mu$m AGN continuum and \hbox{2--10~keV} luminosity constraints for
xFLS ULIRGs. For a given mid-IR luminosity, we typically find a large
decrement between our individual \hbox{X-ray} upper limits and the expected
\hbox{2--10~keV} emission, as estimated from the intrinsic \hbox{X-ray}--to--mid-IR
correlation, demonstrating that all of the xFLS ULIRGs should be
heavily obscured AGN. Only two of the individual 3$\sigma$ upper
limits lie below the conservative \hbox{$N_{\rm H}\sim10^{24}$~cm$^{-2}$}
track, while all but two lie below the inclusive \hbox{$N_{\rm
H}\sim10^{24}$~cm$^{-2} + A_{\rm 6\mu m}\sim2$~mag} track. Thus some
fraction of xFLS ULIRGs are likely to be at least mildly
Compton-thick, but many individual upper limits leave ambiguous
constraints. 

The locations of the two marginally full-band detected xFLS ULIRGs,
MIPS\,8268 and MIPS\,8342, can provide some initial guidance. As noted
in $\S$\ref{data}, neither are formally detected in the
rest-frame \hbox{2--10~keV} sub-band, and thus appear in
Fig.~\ref{fig:lir_lx} as upper limits. The rest-frame \hbox{2--10~keV}
luminosities extrapolated from their observed full-band counts lie
below their upper limits by factors of 2.8 and 1.1, respectively, but
still roughly a factor of 2 above the conservative \hbox{$N_{\rm
H}\sim10^{24}$~cm$^{-2}$} track. Both extrapolations are at least
broadly consistent with our crude band ratio analysis in
$\S$\ref{data}. If these two objects represent the least obscured xFLS
ULIRGs, then we would expect the rest to lie near or below the
conservative track accordingly.

We can improve upon these constraints by stacking the \hbox{X-ray}
counts of the undetected sources in two subsamples, split into crudely
matching redshift ranges (see $\S$\ref{data}). Importantly, we find
that neither sample is statistically detected, with stacked 3$\sigma$
upper limits lying factors of $\approx$1.6 and $\approx$5.2 below the
conservative \hbox{$N_{\rm H}\sim10^{24}$~cm$^{-2}$} track for the
$z\sim1$ and $z\sim2$ samples, respectively. Assuming all sources
contribute roughly equally to the stacked signal, these low limits
imply that $\ga$70\% of the $z\sim1$ and $\ga$90\% of the $z\sim2$
{\it Chandra}-observed xFLS ULIRGs should be at least mildly
Compton-thick at 99\% confidence, and perhaps some fraction even fully
Compton-thick \hbox{($N_{\rm H}\ga10^{25}$~cm$^{-2}$)}. Assuming the
{\it Chandra}-observed objects are representative and follow a normal
distribution, then we can calculate 'margin of error'
constraints\footnote{ When only a subset of the total population is
sampled, the associated error (often termed 'margin of error') on the
fraction of the total population with a particular property can be
calculated using $z \sqrt{p (1-p) \over n} f$, where $z$ is the
'critical value' related to the form of the underlying distribution,
the degrees of freedom, and the confidence interval of interest (in
this case a 1-sided $t$-distribution), $p$ is the fraction of a
particular property found from the subset population, and $f$ is the
finite population correction factor $\sqrt{N-n \over N-1}$, $n$ is the
subset number, and $N$ is the total population
number \citep[e.g.,][]{Peck2008}.} on the potential Compton-thick
fraction for the entire sample. For $z\sim1$, we only stack 3 of 15
total xFLS ULIRGs, resulting in a candidate Compton-thick fraction of
$\ga$25\% at 90\% confidence, but no meaningful constraint at higher
confidences. For $z\sim2$, we stack 11 of 37 total xFLS ULIRGs,
resulting in candidate Compton-thick fractions of $\ga$80\% and
$\ga$70\% for confidences of 90\% and 99\%, respectively. Our general
conclusion that a majority, if not all, xFLS ULIRGs are likely to host
luminous Compton-thick AGN is supported by the fact that the
individual xFLS upper limits lie in the same range as some observed
X-ray luminosities from our IR-bright Compton-thick ULIRG sample
($\S$\ref{brightnature}), implying that the xFLS objects could be even
more extreme.

This result is predicated on the fact that the \hbox{X-ray} decrement
provides a reliable estimate of the obscuration, when compared to the
AGN continuum luminosity at 5.8\,$\mu$m. While we argued in
$\S$\ref{brightnature} that this procedure is robust, emission-line
luminosities from AGN can provide an independent check on the AGN
power estimated from our 5.8\,$\mu$m measurements. Only five xFLS
ULIRGs have robust [O\,{\sc III}] detections among nine observed
(S08), which we plot in the right panel of
Fig.~\ref{fig:lir_lx}. These emission-line luminosities have not been
corrected for contamination by star formation or extinction, primarily
because neither is well quantified. The expected contribution to the
observed $L_{[OIII]}$ from star formation should be
minimal \hbox{($\la$1--10\%)}, as long as the well-established
star-formation correlations between [O\,{\sc III}], [O\,{\sc II}], and
$L_{IR}$ \citep[e.g.,][]{Hopkins2003, Moustakas2006} hold for the xFLS
ULIRGs. As discussed in S08, the effects of internal extinction are
more problematic, since the few sources with direct rest-frame optical
constraints exhibit a wide range of apparent extinctions.

For instance, among the five [O\,{\sc III}]-detected xFLS ULIRGs in
Fig.~\ref{fig:lir_lx}, direct assessment of the optical extinction via
the Balmer decrement is available for four. Assuming no systematic
errors exist between near-IR spectral segments, two sources show no
evidence for extinction, while the other two have $E(B-V)=1.2$ and
$E(B-V)>2.2$, respectively. Among comparable IR-bright,
optically-faint $z\sim2$ sources, \citet{Brand2008} also report
extinctions in the range \hbox{$E(B-V)=1.0$--1.9}. Thus the amount of
optical extinction can vary dramatically from source to source and
potentially be quite severe. There also remains some possibility that
the [O\,{\sc III}] emission does not see the same extinction as the
Balmer lines. We can alternatively estimate $E(B-V)$ based on the
silicate absorption in the mid-IR, which lies in the
range \hbox{$\tau_{9.7}\approx1.3$--2.7} for the five sources and
provides constraints on extinction from the nuclear region itself. The
ratio $A_{\rm V}/\tau_{9.7}$ is principally dependent on the relative
abundances of silicate and graphite grains: $\approx$18 for diffuse
interstellar material (ISM), \hbox{$\approx$18--40} for denser clouds,
and $\approx$9 near the GC \citep[e.g.,][]{Roche1985, Whittet2003,
Chiar2007}. If we conservatively adopt the low GC conversion and
$R_{\rm V}=5$, then we would expect \hbox{$A_{\rm V}\sim12$--24}
or \hbox{$E(B-V)\sim2.3$--4.9}, which is substantially higher than the
direct estimates. Even adopting a somewhat conservative
$E(B-V)\sim1.0$, the correction to [O\,{\sc III}] is already
\hbox{$\sim$1--2~dex}.

The five [O\,{\sc III}]-detected xFLS ULIRGs have \hbox{$\log [ L_{\rm
[OIII]}/\nu L_{\nu} ({\rm 5.8\mu m}) ]$} ratios of -3.0 to -3.5, which
systematically lie \hbox{1--2~dex} lower than the average type~1 and
type~2 AGN/ULIRGs, respectively \citetext{e.g., \citealp{Haas2007};
Bauer et al., in preparation}. The discrepancy is consistent with our
expectation that the [O\,{\sc III}] luminosities are highly
extincted. Thus our uncorrected values should be considered {\it very}
conservative lower limits to the true AGN power. 

To compare against our X-ray limits in the right panel of
Fig.~\ref{fig:lir_lx}, we convert the [O\,{\sc III}] constraints to
X-ray ones using the \hbox{[O\,{\sc III}]-to-X-ray} correlation for
type\,1 AGN
\hbox{($L_{\rm [OIII]}=10^{6.9\pm2.2}L_{\rm 2-10\,keV}^{0.88\pm0.05}$),}
where the emission-line luminosities have not been corrected for
extinction (Bauer et al., in preparation); both luminosities are in
units of erg~s$^{-1}$.\footnote{We note that this \hbox{$L_{\rm
[OIII]}/L_{\rm 2-10\,keV}$} correlation differs significantly from
those found by, e.g., \citet{Netzer2006}, \citet{Panessa2006}
and \citet{Melendez2008}, which primarily appears to be due to the
limited luminosity ranges of the aforementioned samples. For a given
[O\,{\sc III}] luminosity in the high-luminosity, radio-quiet AGN
sample of \citet{Maiolino2007}, for instance, these relations all
overestimate the detected \hbox{X-ray} luminosities by
\hbox{$\sim$1--2~dex}. Importantly, applying any of these other correlations
would result in a significantly larger intrinsic \hbox{X-ray}
luminosity estimate, thus favoring even more \hbox{X-ray} absorption.}
A08, for instance, have recently shown that several
optically-identified Compton-thick AGN also follow the AGN correlation
shown in Fig.~\ref{fig:lir_lx}, when one converts uncorrected-[O\,{\sc
III}] luminosity into \hbox{X-ray} luminosity. We find that the
predicted rest-frame \hbox{2--10~keV} luminosities lie a factor
of \hbox{$\approx$5--6} above the current \hbox{X-ray} limits,
supporting the presence of significant \hbox{X-ray} obscuration. Since
the optical extinction in the xFLS ULIRGs should be much larger than
for the type\,1 AGN used to derive the correlation, these converted
X-ray luminosity constraints should be considered very conservative.

An alternative scenario sometimes invoked is that the hot dust
component is somehow associated with massive, deeply embedded star
formation. Wolf-Rayet and early-type O stars, for instance, are
certainly capable of producing the hard radiation necessary to power
the $L_{\rm 5.8\mu m}$ continua in theory. Strong continuum-dominated
sources like the weak-PAH xFLS ULIRGs, however, lie \hbox{1--2~dex}
above the the \hbox{$L_{\rm 5.8\mu m}$-$L_{\rm PAH(7.7\mu m)}$}
correlation for star-forming galaxies. This effectively rules out
star-formation for all but the most contrived scenarios, but is fully
consistent with known AGN \citep[e.g., S07;][]{Desai2007}. Further
indirect evidence against star formation comes from high-spatial
resolution mid-IR imaging of the most powerful local infrared luminous
AGN, including a few from Table~\ref{tab:brightsample}. Such
observations constrain a large fraction of the mid-IR continuum light
within a region $\la$100~pc in size ($\la$0\farcs3) around the
nucleus, with mid-IR surface brightnesses in excess of
$\approx$10$^{14}$ $L_{\odot}$~kpc$^{-2}$ \citep[e.g.,][]{Soifer2000,
Soifer2003}. While the most extreme 'super' star-clusters known
locally can still generate comparable mid-IR surface
brightnesses \citep[e.g.,][]{Gorjian2001}, the largest sizes of such
star-clusters are only 1--10\,pc, which are much smaller than the
current spatial limits on local ULIRG nuclei. Coupling this 1--2 dex
size deficit with the facts that the xFLS ULIRGs have (1) mid-IR
luminosities 1--2 dex higher than the above local AGN and (2)
relatively weak PAH features, and it seems highly unlikely that super
star-clusters (even exotic ones at high redshift) are capable of
producing the required mid-IR luminosities of order
10$^{12}$--$10^{13}$\,$L_\odot$. As such, AGN appear to be the most
plausible option for powering the bulk of the mid-IR emission in
ULIRGs.

In summary, the strong 5.8\,$\mu$m continuum and [O\,{\sc III}]
luminosities (once corrected for extinction), coupled with the
weak \hbox{X-ray} emission even in stacked \hbox{X-ray} images, argues
for at least mild Compton-thick obscuration in a large fraction of, if
not all, xFLS ULIRGs.

\begin{figure*}
\vspace{-0.1in}
\centerline{
\hglue-0.2cm{\includegraphics[width=14cm]{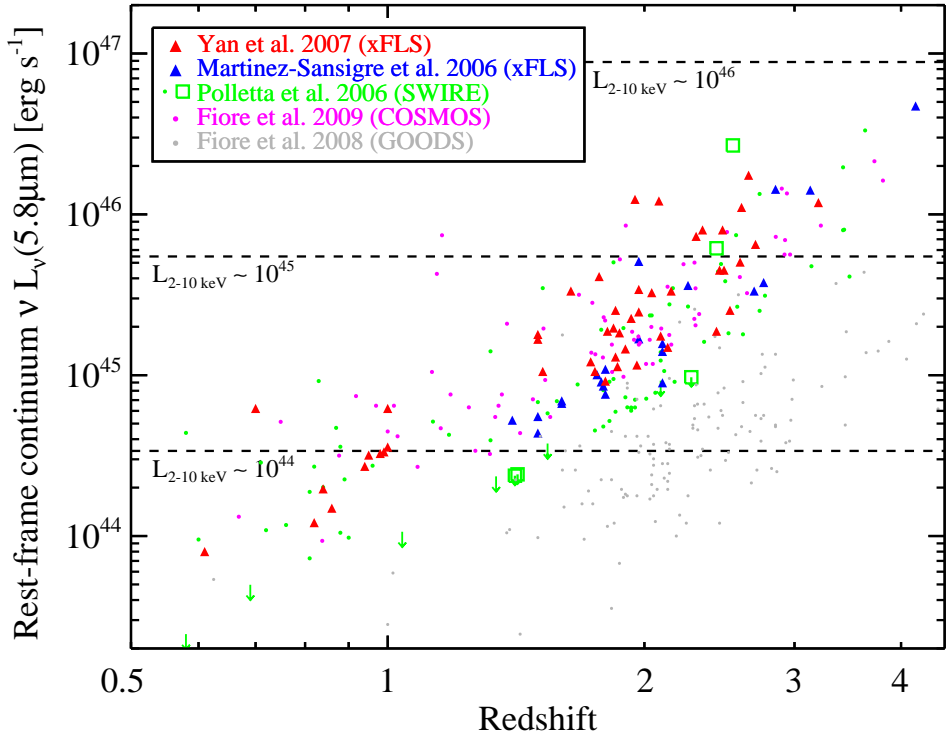}}
}
\vspace{-0.0in} 
\figcaption[L6u_vs_z_color.eps]{
{\footnotesize Rest-frame $L_{\rm 5.8\mu m}$ versus redshift for a select few samples
of candidate Compton-thick AGN including: mid-IR excess objects
studied here (red triangles); \hbox{X-ray} (open green squares) and
IR-selected (green circles) objects from P06; radio-excess objects
from MS06 (blue triangles); mid-IR excess objects from F08/F09 (grey
circles for GOODS selection, magenta circles for COSMOS
selection). Triangles denote samples with robust redshifts ($\ga$90\%
from spectroscopy), while circles denote less-robust samples with
redshifts predominantly from photometry ($\sim$10\% spectroscopy for
P06; $\sim$5\% and $\sim$40\% spectroscopy for F08 GOODS and F09
COSMOS, respectively). Aside from the five \hbox{X-ray} detected
Compton-thick candidates from P06, we only plot the subset of obscured
AGN candidates which are individually \hbox{X-ray} undetected for each
sample, under the assumption that an \hbox{X-ray} detection of any kind
potentially implies something less than Compton-thick
obscuration. Dashed lines denote expected $L_{\rm 2-10~keV}$
luminosities adopting the rest-frame \hbox{5.8\,$\mu$m-to-2--10~keV}
correlation shown in Fig.~\ref{fig:lir_lx}. Rest-frame $L_{\rm 5.8\mu
m}$ values for P06 and MS06 were estimated from interpolation of
best-fit power-law models to the published fluxes, incorporating an
additional boost of $\approx$1.2 to account for apparent spectral
curvature due to extinction; this average boost was determined
empirically from the xFLS ULIRG sample, where rest-frame 5.8\,$\mu$m
continuum fluxes have been determined directly from spectral
decomposition (S07; S08). This estimation method is only appropriate
for sources with mid-IR spectra well-represented by power-law models,
and could deviate somewhat for objects with strong absorption or
emission features. It should also be stressed that while the
Compton-thick AGN nature for the bulk of xFLS ULIRGs appears robust,
the same level of confidence cannot be assumed for the other samples,
for which the relative fractions of sources dominated by
star-formation in the IR or obscured but not Compton-thick are
generally not well established.
\label{fig:l6u_z}}}
\vspace{0.2cm} 
\end{figure*}

\subsection{Comparison to Other Selection Techniques}\label{selection}

There are several different techniques to find mid-IR AGN, and the
common goal of these methods is to identify heavily-obscured, and
potentially Compton-thick AGN that might be missed by \hbox{X-ray},
UV, or radio selection.  These techniques generally fall into a few
overlapping categories: (1) those like Y07 that look for strong hot
dust components relative to stellar continua (``mid-IR excess''), (2)
those that look for hot dust components directly in the
IRAC \hbox{3.6--8.0$\mu$m} bands alone (``mid-IR spectral slope''),
and (3) those that look for strong radio emission above that expected
from star formation (``radio excess''). A recent study
by \citet{Donley2008} has investigated the reliabilities of these
methods.  Given the degree of high quality follow-up data on our
sample, particularly spectroscopic redshifts, and the likelihood that
xFLS ULIRGs host Compton-thick AGN, a lower limit to the number of
potential Compton-thick AGN discovered by other methods can be
assessed by investigating the relative frequency with which xFLS-like
ULIRGs are selected via these other methods. Such comparisons also
allow us to place the xFLS ULIRGs within the context of these other
surveys, particularly in regard to their relative space densities
presented in $\S$\ref{spacedensity}. Table~\ref{tab:select} provides
an overview of our findings, which we discuss in detail below. We
automatically exclude xFLS ULIRGs with unconstrained IRAC colors from
affected comparisons. Also, since sources with direct \hbox{X-ray}
detections are likely to be relatively unobscured, we only make
comparisons between \hbox{X-ray} undetected subsets of the samples.

We begin comparisons with the closely related studies
of \citet[][collectively hereafter MS06]{MartinezSansigre2006,
MartinezSansigre2007} and \citet[][hereafter P06]{Polletta2006}, both
of whom also focus on similarly powerful obscured AGNs (see
Fig.~\ref{fig:l6u_z}). The ``radio-excess'' technique of MS06 selects
only high-redshift ULIRGs (also within the xFLS footprint) which lie
above the radio-to-far-infrared correlation for star-forming
galaxies \citep[e.g.,][]{Condon1992}. This technique targets the
radio-intermediate to radio-loud portion of the obscured QSOs
population, from which the remaining radio-quiet subset can, in
theory, be loosely extrapolated. Within a given redshift bin, the
average 5.8\,$\mu$m luminosity of the MS06 sample is roughly a factor
of two lower, indicating that MS06 samples less powerful
AGN. Importantly, nearly all of the MS06 sample has also been observed
with the {\it Spitzer} IRS \citep{MartinezSansigre2008}, and display
properties very similar to the xFLS ULIRGs, such as strong silicate
absorption \hbox{($\tau_{9.7}$\,$\sim$\,1--2)} and spectra dominated
by strong hot dust continua.  Under their criteria, five of the xFLS
ULIRGs sources should nominally have been selected, although three of
these appear to have been rejected from that sample due to
source-blending issues in the IRAC 3.6$\mu$m band. The fraction of
xFLS ULIRGs selected by this method thus lies somewhere between
\hbox{5--10\%}. While this seems low, extrapolating from the fraction of
radio-intermediate AGN to the full population yields a more
substantial overlap.

\begin{deluxetable*}{lcc}
\tabletypesize{\scriptsize}
\tablewidth{0pt}
\tablecaption{Selection Technique Comparison\label{tab:select}}
\tablehead{
\colhead{Study} & 
\colhead{Selection Criteria} & 
\colhead{xFLS Recovery}
}
\tableheadfrac{0.05}
\startdata
\citet{Yan2007} & $S_{24{\rm \mu m}} > 0.9\,{\rm mJy}$, $\alpha(24,8.0) > 0.5$, $\alpha(24,R) > 1.0$ & --- \\
\citet{Lacy2004} & $S_{24{\rm \mu m}} > 4.4\,{\rm mJy}$ or $S_{8.0{\rm \mu m}} > 1\,{\rm mJy}$, $r(5.8,3.6) >-0.1$, $r(8.0,4.5) > -0.2$, $r(8.0,4.5) \le 0.8[ r(5.8,3.6) ] + 0.5$ & $\approx$65\%* \\
\citet{Stern2005} & $[R] \le 21.5$, $S_{24{\rm \mu m}} > 1.0\,{\rm mJy}$, $([5.8]-[8.0])>-0.07$, & $\approx$30\%* \\
                  & $([3.6]-[4.5])>0.2([5.8]-[8.0])-0.16$, $([3.6]-[4.5])>2.5([5.8]-[8.0])-2.3$ &  \\
\citet{Cool2006} & $[I] \le 22$, $S_{24{\rm \mu m}} > 1.0\,{\rm mJy}$, $([3.6]-[4.5])>-0.1$, $([5.8]-[8.0])>-0.05$ & $\approx$100\%* \\
\citet{Donley2007} & $S_{24{\rm \mu m}} > 80\,{\rm \mu Jy}$, $\alpha_{IRAC} \le -0.5$, where $f_{\nu}\propto \nu^{\alpha}$, $P>0.1$ & $\approx$70\% \\
\citet{Polletta2006} & monotonic flux increase in $\ge$ 3 mid-IR bands, & $\sim$70\% \\
                     & $\alpha_{2--24\mu m} \le -1$, where $f_{\nu}\propto \nu^{\alpha}$, $\chi_{\nu}<13.2(-\alpha_{2--24\mu m}-1) \le 20$, & \\
                     & $r(3.6,g')\ge1.18$, $r(3.6,r')\ge1.11$, $r(3.6,i')\ge1.0$, SED fitting & \\
\citet{MartinezSansigre2006} & $S_{24{\rm \mu m}} > 0.3\,{\rm mJy}$, $S_{3.6{\rm \mu m}} \le 45\,{\rm \mu Jy}$, $0.35\,{\rm mJy} \le S_{1.5{\rm GHz}} \le 2\,{\rm mJy}$ & $\approx$5--10\% \\ 
\citet{Fiore2009} & $S_{24{\rm \mu m}} > 0.55\,{\rm mJy}$ (COSMOS) or $40\,{\rm \mu Jy}$ (GOODS), $r(24,R) > 3.0$, $[R]-[K]>4.5$ & $\approx$40\% \\
\citet{Dey2008} & $S_{24{\rm \mu m}} > 0.3\,{\rm mJy}$, $r(24,R) > 3.0$ & $\approx$50\% \\
\enddata
\tablecomments{
{\it Col. (1)} Selection technique.
{\it Col. (2)} Flux/magnitude cutoffs and color selection criteria. Here 
$\alpha(\lambda_{1},\lambda_{2})=\log ({\nu f_{\nu}(\lambda_{1})\over \nu f_{\nu}(\lambda_{2})})$ and 
$r(\lambda_{1},\lambda_{2})=\log ({f_{\nu}(\lambda_{1})\over f_{\nu}(\lambda_{2})})$. 
Brackets ``$[\lambda]$'' denote AB magnitudes.  $\lambda$ itself
denotes the {\it Spitzer} IRAC and MIPS band (given as central
wavelength in units of $\mu$m) or the standard optical/near-IR
band.
{\it Col. (3)} Fraction of sources selected by the xFLS ULIRGs
criteria which would also be selected by another technique. The
fractions denoted with a '*' have been assessed without the stated
$S_{24{\rm \mu m}}$ selection criteria, since these would typically
reject the bulk of xFLS ULIRGs outright.
}
\end{deluxetable*}

P06, on the other hand, use a number of criteria (both ``mid-IR excess'' and
``mid-IR spectral slope'') to remove sources with non-power-law SEDs from
their IR-selected sample (see Table~\ref{tab:select}), selecting
sources with a wide range of SED-types, luminosities, and obscuration
levels. They find that $\approx$35\% of their sample are \hbox{X-ray}
detected, with only two \hbox{X-ray}-detected sources showing evidence for
\hbox{$N_{\rm H}>10^{24}$~cm$^{-2}$} absorption. The absorption for the \hbox{X-ray}
undetected subset of their sample should typically be comparable to or
larger than the \hbox{X-ray} detected sources, but this absorption
distribution has yet to be firmly established. The 5.8\,$\mu$m
luminosity and redshift range of the xFLS sources matches the upper
end of the P06 sample, and $\sim$70\% of xFLS ULIRGs should be
selected by the P06 criteria. The large overlap suggests that a
sizable number of Compton-thick AGN likely exist within the P06
sample. How this extends to the lower luminosity sources, however, and
particularly to the unusually large number that lie near the flux
limit of their survey, is unclear.

Applying the most commonly used IRAC AGN color-selection cuts
from \citet{Lacy2004} and \citet{Stern2005}, we find that
$\approx$70\% and $\approx$30\% of the xFLS ULIRGs are selected,
respectively. The low percentage for the latter stems from the fact
that the majority of xFLS ULIRGs lie immediately to the right of
the \cite{Stern2005} region. Extending the \citet{Stern2005} region
for $z>1$ objects as \citet{Cool2006} have done, for example, recovers
an impressive $\approx$100\% of our sample. A major limitation in all
of the above studies, however, is the adoption of either shallow
optical and/or mid-IR flux cutoffs, which effectively remove all xFLS
ULIRGs regardless of color-selection criteria. In the case
of \citet{Lacy2004}, the factor of $\approx$5 higher 24$\mu$m cutoff,
coupled with the expected strongly declining space density evolution
for ULIRGs, is likely to severely limit the number of comparably
luminous but more nearby xFLS-like ULIRGs detected.
For \cite{Stern2005} and \citet{Cool2006}, the bright optical
selection will effectively remove a large fraction of highly-obscured
ULIRGs like those studied here. While extending the above techniques
to deeper samples would help select ULIRGs like those in the xFLS,
doing so could be counterproductive since it could also introduce many
new star-forming contaminants \citep[see][]{Donley2008}. Adding an
additional color selection to remove $z<1.2$ galaxies, such as an
$R-K$ criteria, could effectively address this apparent limitation.

The other main ``mid-IR spectral slope'' technique used extensively is
the power-law galaxy (PLG) criteria of \citet{Alonso-Herrero2006}
and \citet{Donley2007}, which \citet{Donley2008} contend recovers the
majority of high-quality AGN candidates. All of the xFLS ULIRGs would
qualify as PLGs based on their {\it underlying} mid-IR AGN continuum
slopes as found from spectral decomposition. However, using only the
observed IRAC data alone as stipulated in \citet{Donley2007}, only
$\approx$70\% of the xFLS ULIRGs are recovered as PLGs. While not in
direct contrast with
\citet{Donley2008}, our result implies that there exists a
rather substantial population of legitimate mid-IR excess AGN which
are currently missed by the PLG technique. These non-PLG objects
typically either had poor IRAC constraints (and subsequently large
uncertainties in their spectral slopes) or were dominated by stellar
continua well into the IRAC bands but had strong hot-dust components
at longer wavelengths (see S07). Intriguingly, several
of the PLG-selected xFLS ULIRGs also had relatively strong host
stellar continuum components, demonstrating that the PLG method does
select some composite AGN.

Among the ``mid-IR excess'' techniques, the \citet[][hereafter F08 and
F09]{Fiore2008, Fiore2009} and \citet{Dey2008} selection criteria are
the most closely related to Y07, although both employ much lower
24$\mu$m flux cutoffs, have no 24$\mu$m-to-8$\mu$m flux selection, and
require a factor of $\approx$3 larger 24$\mu$m to R-band
excess. F08/F09 additionally implement an $R-K$ cut to remove
low-redshift contamination from star-forming galaxies. The lower flux
cutoff there probes deeper into the overall 24$\mu$m source population
to select both lower luminosity and higher redshift objects (see
Fig.~\ref{fig:l6u_z}), while the larger mid-IR excess picks only the
reddest ULIRGs. The lack of 24$\mu$m-to-8$\mu$m selection, however,
should lead to the inclusion of many sources with either less
obscuration and/or hotter dust continua. We find that only
$\approx$40\% and $\approx$50\% of xFLS ULIRGs lie at the extremes
selected by the F08/F09 and \citet{Dey2008} techniques,
respectively. There is a mild tendency for these techniques to
predominantly select higher redshift sources as expected, although
several high-z xFLS ULIRGs are still notably excluded. If all of the
5.8$\mu$m luminosity in F08/F09-selected objects stems from obscured
AGN activity, then the \hbox{X-ray} stacking detection performed by
these authors equates to an average mid-IR/\hbox{X-ray} decrement of
$\approx800$, which is an order of magnitude below our
xFLS \hbox{X-ray} luminosity stacking upper limits, and hence fully
consistent with our results. Another mid-IR excess selection method is
that of \citet[][hereafter D07]{Daddi2007b}, which selects a large
number of high-redshift, moderate-luminosity, obscured AGN candidates
comparable to the GOODS sources in F08. This selection technique is
too complex to employ here, since it relies on assumption-dependent
excesses between the dust-corrected UV and IR star formation rates in
each object. Nonetheless we note that the D07 sample has an average
5.8$\mu$m excess luminosity of $\sim$3$\times10^{44}$~erg~s$^{-1}$,
which is comparable to those of our $z$$\sim$1 objects, and an X-ray
stacking detection an order of magnitude below our $z\sim1$ X-ray
luminosity upper limits, which again is fully compatible with our
results.

While it is rare to find full consistency between Y07 and other
selection methods, there appears to be enough overlap (if we neglect
shallow flux criteria) to demonstrate that many of these methods
select a substantial subset of powerful, potentially Compton-thick
ULIRGs like the ones characterized here. Aside from MS06, however, a
fundamental limitation of these other studies is their reliance on
broad-band photometry and, in several cases, photometric
redshifts. This critical lack of precise redshifts and
spectrally-deconvolved SEDs makes it impossible to distinguish
unambiguously between mid-IR excesses due to hot dust from AGN and
those from strong PAH features due to vigorous star formation. While
this may be a relatively small concern at high mid-IR luminosities
where AGN are thought to dominate (e.g., \hbox{$L_{\rm 5.8\mu
m}$$\ga$$10^{45}$ erg~s$^{-1}$}), it is likely to cause severe
problems at lower mid-IR luminosities where star formation is expected
to be ubiquitous \citep[e.g., S07;][]{Lacy2007, MartinezSansigre2008}.

\begin{figure}
\vspace{-0.1in}
\centerline{
\hglue-0.5cm{\includegraphics[width=9.8cm]{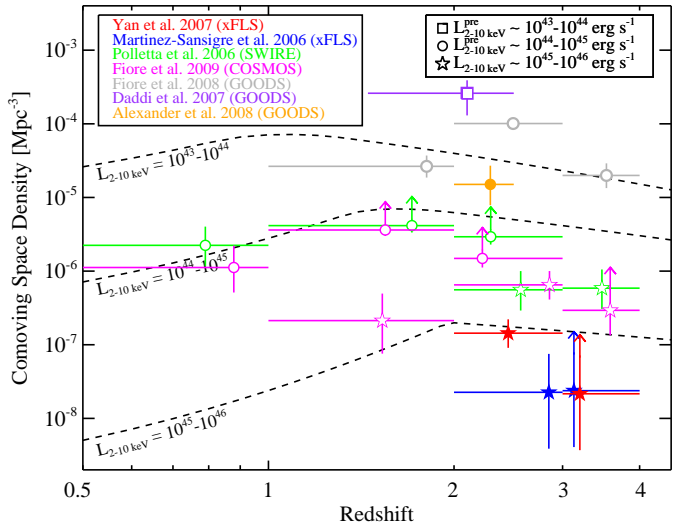}}
}
\vspace{-0.0in} 
\figcaption[CSD_vs_z_color.eps]{
{\footnotesize Co-moving space density of candidate Compton-thick AGN in three
approximate rest-frame, intrinsic \hbox{X-ray} luminosity ranges:
\hbox{$L_{2-10\,keV}=10^{43}$--10$^{44}$ erg~s$^{-1}$} (squares),
\hbox{$L_{2-10\,keV}=10^{44}$--10$^{45}$ erg~s$^{-1}$} (circles), and
\hbox{$L_{2-10\,keV}=10^{45}$--10$^{46}$ erg~s$^{-1}$} (stars). Only a subset
of constraints from Table~\ref{tab:csd} are shown. The color-coding of
the samples is identical to Fig.~\ref{fig:l6u_z}, with the additions
of Compton-thick candidate AGN within the GOODS region from D07
(purple) and A08 (orange). Similar to Fig.~\ref{fig:l6u_z}, samples
with mostly robust spectroscopic redshifts are shown as filled
symbols, while less-robust photometrically-constrained samples are
shown with open symbols. For comparison, we show the predicted space
densities of Compton-thick AGNs from the G07 CXRB model for these
three \hbox{X-ray} luminosity bins (dashed lines). The number of
candidate Compton-thick xFLS ULIRGs are roughly equal to the predicted
values to within errors in the
\hbox{$L_{2-10\,keV}=10^{44}$--10$^{45}$ erg~s$^{-1}$} bin, while P06 and F09
lie a factor of \hbox{$\sim$3--4} higher. The fact that these sample
selection techniques do not fully overlap implies that Compton-thick
QSOs are likely ubiquitous at high redshift.
\label{fig:csd_z}}}
\vspace{0.2cm} 
\end{figure} 

\subsection{The Space Density of Heavily-Obscured Accretion}\label{spacedensity}

We now investigate the space densities of the candidate Compton-thick
AGN samples discussed in $\S$\ref{selection} in several redshift and
predicted intrinsic \hbox{X-ray} luminosity bins. Approximate values
of intrinsic rest-frame $L_{\rm 2-10 keV}$ were predicted from
rest-frame $L_{\rm 5.8\mu m}$ assuming the conversion
above.\footnote{The choice of correlation slope and intercept do not
have a strong effect on the estimated luminosities (and there space
densities), and lead to variations of perhaps $\approx$2 in either
direction. Additionally, correcting rest-frame $L_{\rm 5.8\mu m}$ for
extinction could raise it a factor of $\approx$2 on average, implying
a similar increase to intrinsic rest-frame $L_{\rm 2-10 keV}$.} A key
concern is that although these samples all span a wide range of
luminosities and redshifts (see Fig.~\ref{fig:l6u_z}), their selection
is often potentially biased or incomplete within a particular bin due
to limited solid angle coverage, restrictive selection criteria, or
survey flux limits. Table~\ref{tab:csd} provides space density
estimates for the full range of parameter space probed, while
Fig.~\ref{fig:csd_z} highlights the best constraints for each of the
samples. For bins which are obviously incomplete, we assume the
derived values are lower limits. We again strongly caution over
interpretation of sources in the comparison samples which often lack
both complete optical/near-IR spectroscopic identification and
well-sampled SEDs with IRS spectroscopy to decouple possible emission
mechanisms. Contamination from star formation, particularly at lower
mid-IR luminosities (as discussed in $\S$\ref{selection}), could be
substantial. Since this contamination is not yet well-constrained, we
liberally place these comparison samples at their highest space
densities using all candidate objects.

\begin{deluxetable*}{lcccc}
\tabletypesize{\scriptsize}
\tablewidth{0pt}
\tablecaption{Compton-Thick Candidate Space Densities\label{tab:csd}}
\tablehead{
\colhead{Sample} & 
\colhead{Area} &
\colhead{$z$} &
\colhead{$L_{2-10\,keV}=10^{44}$--10$^{45}$ erg~s$^{-1}$} & 
\colhead{$L_{2-10\,keV}=10^{45}$--10$^{46}$ erg~s$^{-1}$}
}
\tableheadfrac{0.05}
\startdata
\citealp{Yan2007} (xFLS) & 3.26* & 0.5--1.0 & \llap{$<$}1.44E$-$07\rlap{ ( 0)}                   & \llap{$<$}1.44E$-$07\rlap{ (0)} \\
           &       & 1.0--2.0 & \llap{$>$}1.32$^{+0.39}_{-0.31}$E$-$07\rlap{ (18)} & \llap{$<$}1.64E$-$07\rlap{ (0)} \\
           &       & 2.0--3.0 & \llap{$>$}1.85$^{+0.74}_{-0.53}$E$-$07\rlap{ ( 9)} &           1.44$^{+0.68}_{-0.46}$E$-$07\rlap{ (7)}\\
           &       & 3.0--4.0 & ---                                                & \llap{$>$}2.16$^{+4.39}_{-1.56}$E$-$08\rlap{ (1)}\\
\hline
\citealp{Yan2007} (xFLS) & 3.26* & 0.5--1.0 & \llap{$<$}1.44E$-$07\rlap{ ( 0)}                   & \llap{$<$}1.44E$-$07\rlap{ (0)} \\
w/o strong PAH sources   &       & 1.0--2.0 & \llap{$>$}9.53$^{+3.45}_{-2.60}$E$-$08\rlap{ (13)} & \llap{$<$}1.64E$-$07\rlap{ (0)} \\
                         &      & 2.0--3.0 & \llap{$>$}1.44$^{0.68}_{-0.46}$E$-$07\rlap{ ( 7)} &        1.03$^{+0.61}_{-0.39}$E$-$07\rlap{ (5)}\\
           &       & 3.0--4.0 & ---                                                & \llap{$>$}2.16$^{+6.27}_{-2.24}$E$-$08\rlap{ (1)}\\
\hline
\citealp{MartinezSansigre2006} (xFLS) & 3.70 & 0.5--1.0 & \llap{$<$}5.08E$-$07\rlap{ ( 0)}                   & \llap{$<$}5.08E$-$07\rlap{ (0)} \\
           &       & 1.0--2.0 & \llap{$>$}3.10$^{+1.18}_{-0.88}$E$-$07\rlap{ (12)} & \llap{$<$}1.45E$-$07\rlap{ (0)} \\
           &       & 2.0--3.0 & \llap{$>$}1.36$^{+0.81}_{-0.54}$E$-$07\rlap{ ( 6)} &           2.26$^{+5.25}_{-1.87}$E$-$08\rlap{ (1)}\\
           &       & 3.0--4.0 & ---                                                & \llap{$>$}2.38$^{+5.52}_{-1.97}$E$-$08\rlap{ (1)}\\
\hline
\citealp{Polletta2006} (SWIRE) & 0.60 & 0.5--1.0 &           2.24$^{+1.78}_{-1.07}$E$-$06\rlap{ ( 4)} & \llap{$<$}3.13E$-$06\rlap{ (0)} \\
           &       & 1.0--2.0 & \llap{$>$}4.15$^{+0.98}_{-0.81}$E$-$06\rlap{ (26)} & \llap{$<$}8.93E$-$07\rlap{ (0)} \\
           &       & 2.0--3.0 & \llap{$>$}2.93$^{+0.79}_{-0.63}$E$-$06\rlap{ (21)} & 5.58$^{+4.43}_{-2.66}$E$-$07\rlap{ (4)}\\
           &       & 3.0--4.0 & \llap{$>$}2.93$^{+3.90}_{-1.89}$E$-$07\rlap{ ( 2)} & 5.87$^{+4.66}_{-2.80}$E$-$07\rlap{ (4)}\\
\hline
\citealp{Fiore2009} (COSMOS) & 0.90 & 0.5--1.0 &          1.12$^{+1.10}_{-0.61}$E$-$06\rlap{ ( 3)} & \llap{$<$}2.09E$-$06\rlap{ (0)} \\
           &       & 1.0--2.0 & \llap{$>$}3.62$^{+0.73}_{-0.62}$E$-$06\rlap{ (34)} &           2.13$^{+2.83}_{-1.37}$E$-$07\rlap{ (2)}\\
           &       & 2.0--3.0 & \llap{$>$}1.49$^{+0.47}_{-0.37}$E$-$06\rlap{ (16)} &           6.50$^{+3.52}_{-2.39}$E$-$07\rlap{ (7)}\\
           &       & 3.0--4.0 & ---                                                & \llap{$>$}2.93$^{+2.87}_{-1.59}$E$-$07\rlap{ (3)}\\
\hline
\citealp{Fiore2008} (GOODS-S) & 0.04 & 0.5--1.0 & \llap{$<$}4.73E$-$05\rlap{ ( 0)}                   & \llap{$<$}4.73E$-$05\rlap{ (0)} \\
           &       & 1.0--2.0 &           2.65$^{+1.07}_{-0.79}$E$-$05\rlap{ (11)} & \llap{$<$}1.35E$-$05\rlap{ (0)} \\
           &       & 2.0--3.0 &           1.01$^{+0.17}_{-0.15}$E$-$04\rlap{ (48)} & \llap{$<$}1.18E$-$05\rlap{ (0)} \\
           &       & 3.0--4.0 &           1.99$^{+0.91}_{-0.65}$E$-$05\rlap{ ( 9)} & \llap{$<$}1.24E$-$05\rlap{ (0)} \\
\hline
\citealp{Alexander2008} (GOODS-N) & 0.04  & 2.0--2.5 &          1.50$^{+1.20}_{-0.72}$E$-$05\rlap{ ( 4)} & --- 
\enddata
\tablecomments{
{\it Col. (1)} Sample.
{\it Col. (2)} Area of sample, in units of deg$^{-1}$. '*' --- Note for our
sample, the original area (3.7 deg$^{-1}$) has been reduced by the
fraction of original sources selected by the Y07 criteria
(59) but not followed-up with {\it Spitzer} IRS; see Y07
for further details.
{\it Col. (3)} Redshift range.
{\it Cols. (4--5)} Space densities derived from number of sources (in
parenthesis) in the rest-frame \hbox{2--10~keV} luminosity ranges of
$10^{44}$-$10^{45}$~erg~s$^{-1}$ and $10^{45}$-$10^{46}$~erg~s$^{-1}$,
respectively, as shown in Fig.~\ref{fig:l6u_z}. Errors quoted are
based on counting statistics only assuming \citet{Gehrels1986}, and
are quoted at 1$\sigma$ confidence when a source exists within a bin
and 3$\sigma$ for upper limits. Bins shown as lower limits suffer from
incompleteness due to flux limits of the various surveys, as can be
seen in Fig.~\ref{fig:l6u_z}.}
\end{deluxetable*}

For comparison, we also show in Fig.~\ref{fig:csd_z} the CXRB
synthesis predictions from G07 for three different luminosity ranges
of Compton-thick AGN. These curves should be regarded as approximate,
since considerable uncertainty remains in the redshift, luminosity,
and column density distributions of the sources that comprise the full
CXRB. In particular, high-luminosity AGN such as these considered here
are predicted to contribute relatively little power to the overall
CXRB, and conversely cannot be strongly constrained by such models. We
note that G07 adopts a model for intrinsic rest-frame AGN
with \hbox{$L_{\rm 2-10~keV}$$\ga$10$^{44}$~ergs~s$^{-1}$} consisting
of equal parts unobscured \hbox{($N_{\rm H}\la10^{22}$~cm$^{-2}$)},
obscured Compton-thin \hbox{($N_{\rm
H}\approx10^{22}$--$10^{24}$~cm$^{-2}$)}, and Compton-thick AGN. They
consider all AGN with $N_{\rm H}\ge10^{22}$~cm$^{-2}$ (both
Compton-thin and Compton-thick) as obscured and $N_{\rm
H}<10^{22}$~cm$^{-2}$ as unobscured, such that their model implies an
expected obscured-to-unobscured ratio (hereafter simply ``obscured
fraction'') of 2:1 among luminous AGN with the G07 model. We stress
that the space densities of unobscured
QSOs \citep[e.g.,][]{Hasinger2005, Hopkins2007} and Compton-thin
QSOs \citep[e.g.,][]{Ueda2003, LaFranca2005, Silverman2008} are
well-constrained by observations, while the number of Compton-thick
AGN have only been inferred from modeling. We investigate here how
many Compton-thick AGN might be expected from several mid-IR selected
surveys. Because these three classes of AGN are equal in the G07
model, the lowest two dashed Compton-thick AGN curves also represent
the expected space densities of unobscured and obscured Compton-thin
QSOs; as such, the lines serve as useful visual benchmarks to assess
the potential obscured fractions among various samples.

Considering first the AGN between predicted \hbox{$L^{\rm pre}_{\rm
2-10~keV}$$\approx$10$^{45}$--10$^{46}$~ergs~s$^{-1}$}, we find that
the xFLS ULIRG sample yields a space density
of \hbox{$\Phi\approx$(1.44$^{+0.68}_{-0.46}$)$\times10^{-7}$~Mpc$^{-3}$}
at \hbox{$z=2$--3}, where the xFLS selection technique was designed to
be most sensitive. This value is already $\approx$90\% of that
predicted by the G07 model. However, as determined in
$\S$\ref{xflsnature} under the ``conservative'' \hbox{$N_{\rm
H}\ga10^{24}$~cm$^{-2}$} selection criteria, assuming the IRS-derived
5.8\,$\mu$m AGN continuum luminosities provide a good proxy for the
expected \hbox{X-ray} luminosity, our subset sampling statistics only
allow us to cautiously assume that $\ga$25\% of the $0.5<z<2.0$ and
$\ga$80\% of $2.0<z<4.0$ xFLS ULIRGs here are Compton-thick at 90\%
confidence. Assuming equal parts unobscured and Compton-thin AGN, the
above constraint on the Compton-thick space density becomes $\ga$70\%
that of G07, or equivalently an obscured fraction of $\ga$$1.7$:1
among powerful \hbox{$L^{\rm pre}_{\rm
2-10~keV}$$\approx$10$^{45}$--10$^{46}$~ergs~s$^{-1}$} QSOs. Our other
high-luminosity redshift bin is a factor of a few lower, but remains
consistent with the \hbox{$z=2$--3} prediction due to potential
incompleteness. In the lower range of \hbox{$L^{\rm pre}_{\rm
2-10~keV}$$\approx$10$^{44}$--10$^{45}$~ergs~s$^{-1}$}, however, the
xFLS detects only $\sim$10\% of the objects predicted by G07,
illustrating that the 24$\mu$m flux limit is far too shallow to detect
the majority of these potential faint, high-z ULIRGs; note that
the \hbox{$L^{\rm pre}_{\rm
2-10~keV}$$\approx$10$^{44}$--10$^{45}$~ergs~s$^{-1}$} values from the
xFLS and MS06 samples have not been plotted in Fig.~\ref{fig:csd_z} to
reduce visual clutter, but are listed in Table~\ref{tab:csd} for
completeness.

As shown in Table~\ref{tab:csd}, excluding strong-PAH ULIRGs where the
$L_{\rm 5.8\mu m}$ continuum measurements are not as well-constrained
does not strongly affect our results. For instance, even if we
pessimistically neglect the two strong-PAH sources in the $z=2$--3
bin, the obscured fraction is still $\ga$$1.5$:1. We caution that
these estimates reflect counting and sampling statistics for the xFLS
sample only, and do not account for systematic errors such as the
slope or dispersion in the $L_{\rm 5.8\mu m}$-$L_{\rm 2-10~keV}$
conversion, the unknown mid-IR extinction corrections, or sample
selection completeness for Compton-thick AGN. As we have argued in
previous sections, these effects largely skew toward {\it
substantially underestimating} the true number of Compton-thick AGN
and the $L_{\rm 5.8\mu m}$ values of such AGN, ensuring that our
derived obscured fraction limits should be relatively robust. As such,
our results imply that the simple mid-IR selection criterion laid out
in Y07 can be highly efficient for finding large numbers of powerful
Compton-thick QSOs candidates. To pick up more modest (and typical)
Compton-thick QSOs, however, we must move to deeper mid-IR surveys.

Looking at the estimates from the P06 and F09 techniques, both appear
to find exceptionally high numbers of obscured QSO candidates in
several redshift and predicted luminosity ranges. At \hbox{$L^{\rm
pre}_{\rm 2-10~keV}$$\approx$10$^{45}$--10$^{46}$~ergs~s$^{-1}$},
these estimates lie a factor of $\approx$3 above the G07 curve,
implying a huge population of heavily-obscured AGN and an obscured
fraction of perhaps $\ga$4:1. Even in the \hbox{$L^{\rm pre}_{\rm
2-10~keV}$$\approx$10$^{44}$--10$^{45}$~ergs~s$^{-1}$} regime, where
these surveys become incomplete, these studies still find space
density lower limits at roughly comparable levels to the G07
predictions. Intriguingly, these constraints for both luminosity
ranges are consistently high across several adjacent redshift bins,
indirectly arguing against strong contamination by wide PAH features,
which we might expect to dominate primarily around $z\sim2$ when
rest-frame PAH lines pass through the 24$\mu$m bandpass.

Among the deeper pencil-beam surveys, we show the results of A08 and
F08 in the GOODS regions for \hbox{$L^{\rm pre}_{\rm
2-10~keV}$$\approx$10$^{44}$--10$^{45}$~ergs~s$^{-1}$} AGN. The A08
sample consists of four relatively secure Compton-thick AGN candidates
in the GOODS-N region at \hbox{$z=2$--2.5}, all of which have robust
optical/mid-IR spectroscopy and deep \hbox{X-ray} constraints. The
space density of such sources is a factor of $\approx$2 higher than
the G07 model curve and implies an obscured fraction of perhaps
$\approx$3:1 overall. The F08 sample, on the other hand, employs the
same method as F09 but to substantially fainter 24$\mu$m fluxes. As
such, they find Compton-thick AGN candidate space densities as high as
a factor of \hbox{$\sim$4--20} above the G07 predictions, implying an
fraction well in excess of $\approx$5:1. In contrast to the
consistency across redshift bins for the \hbox{$L^{\rm pre}_{\rm
2-10~keV}$$\approx$10$^{45}$--10$^{46}$~ergs~s$^{-1}$} sources
selected by this method, however, the \hbox{$z=2$--3} bin here is a
factor of 5 higher than adjacent bins. While this could be a
legitimate evolutionary trend, such a high outlier arouses suspicion
and may perhaps best be explained by contamination from star-forming
PAH features \citep[e.g.,][]{Murphy2009b}.\footnote{ This many
heavily-obscured AGN could also be difficult to reconcile with limits
imposed by the local black hole mass density \citep[e.g.,][]{Yu2002,
Marconi2004, Merloni2004, LaFranca2005, Shankar2007}.} For comparison
we also show the constraints of D07, who probe slightly lower AGN
luminosities than F08, in the range \hbox{$L^{\rm pre}_{\rm
2-10~keV}$$\approx$10$^{43}$--10$^{44}$~ergs~s$^{-1}$} over the
redshift range $z$$=$1.4--2.5. Their space density constraint is
similarly high, with no data to compare against at higher or lower
redshifts. Like F08, this dataset also presumably suffers from
uncertainties both in AGN luminosity determinations and in the
fraction of the sample which contribute to the AGN signal. The
disparity of the F08 and D07 results in the $z\sim2$ regime highlights
the need for deep mid-IR spectroscopy to dispel the current ambiguity
among the fainter candidate Compton-thick AGN population.

In general, it appears that the relatively conservative xFLS criteria
selects a substantial number of strong Compton-thick AGN candidates
which are consistent with current predictions, but should be
considered by no means exhaustive. Other techniques such as those
adopted by F09 or P06 could present immensely effective methods for
identifying obscured AGN population, at least for high-luminosity
samples. Clearly substantial caution must be exercised until
individual AGN and star formation contributions can be effectively
pinned down in these samples. Nonetheless, even if only a minority of
the candidates are eventually confirmed, the sheer number of luminous
Compton-thick AGN candidates found by all of the various selection
techniques, which only partially overlap, argues in favor of much
higher obscured fractions at these extreme AGN luminosities than have
been determined from \hbox{X-ray}-detected studies alone. Such
fractions could be well in excess of 4:1 if these various techniques
indeed select a high percentage of secure Compton-thick AGN. Where
this leaves recent claims about the luminosity-dependence of
obscuration \citep[e.g.,][]{Hasinger2008} is unclear.

\section{Conclusion}\label{conclude}

The {\it Chandra} \hbox{X-ray} upper limits we report here indicate
that the substantial obscured AGN accretion found among mid-IR
selected, high-redshift ULIRGs in the xFLS is likely to be
Compton-thick in nature. The number density of these candidate
Compton-thick AGN already nearly equals that of unobscured QSOs and
obscured Compton-thin AGN \citetext{e.g., such as the comparably
numerous AGN studied by \citealp{Weedman2006}
and \citealp{Brand2008}}, suggesting an obscured fraction of at least
$\approx$$1.7$:1. We regard this constraint as a robust
lower limit, since the assumptions made to achieve this result tend to
substantially underestimate 1) the true number of Compton-thick AGN
since our mid-IR selection only partially overlaps with candidate
Compton-thick AGN selected by other techniques, and 2) the $L_{\rm
5.8\mu m}$ values of the AGN we do select. This fraction could easily
exceed 4:1 among QSO-luminosity AGN if even a fraction of the
candidates from various selection methods are confirmed by mid-IR
spectroscopy.

We stress that the uniqueness of our sample, with its relatively
simple selection criteria, complete Spitzer-IRS spectroscopy, and
UV-to-radio SEDs, provides several independent lines of evidences for
powerful AGN activity, and should be regarded as far more reliable
than similar photometric-based samples. This is a major concern
particularly for fainter mid-IR samples (e.g., D07; F08) which attempt
to tease out small AGN contributions from large samples of objects
with dominant starburst components. Unfortunately, the necessary
mid-IR spectroscopic identification and deconvolution required to
confirm these controversial faint sample results must await the launch
of the {\it James Webb Space Telescope}.

Such large obscured AGN fractions at high luminosities have a few
important implications. First, this result adds uncertainty to recent
optical and \hbox{X-ray} constraints that call for some
luminosity-dependence in obscuration and/or the opening angle of the
putative AGN torus \citep[so-called ``receding torus'' models,
e.g.,][]{Lawrence1991, Hasinger2005, Maiolino2007,
Hasinger2008}. Whether this apparent discrepancy in number density
ultimately stems from a short (but obviously ubiquitous) evolutionary
stage of obscured growth or a more fundamental, long-standing property
of the immediate AGN environment remains to be seen. If similar
obscured fractions can be confirmed over a larger range of redshifts
and luminosities, it would certainly place significantly stronger
constraints on the form and nature of SMBH evolution, particularly in
relation to satisfying both the local bulge luminosity and CXRB
constraints. Second, the implied number of such powerful, dust-laden
AGN may provide a plentiful driver for AGN feedback
scenarios \citep[e.g.,][]{Silk1998, Haehnelt1998, Fabian1999,
Murray2005, Fabian2008} and may ultimately lead to considerable
enrichment of intergalactic material \citep[e.g.,][]{Dave2001,
Adelberger2003, Menard2009}.

\acknowledgements
This work is based on observations made with the {\it Spitzer} Space
Telescope, which is operated by the Jet Propulsion Laboratory and
Caltech under contract with NASA, and the {\it Chandra} \hbox{X-ray}
Observatory, which is operated by the Smithsonian Astrophysical
Observatory under contract with NASA.
We thank R. Chary and R. Mushotsky for useful discussions, and F. Fiore
for providing data on the COSMOS and GOODS sources.
Finally, we thank the anonymous referee for comments that improved the
content and presentation of the paper.
We gratefully acknowledge the financial support of {\it Chandra}
awards PF4-50032 (FEB), GO7-8106X (LY) and GO9-0134C (FEB), the Royal
Society (DMA), and the Leverhulme Trust (DMA).

{\it Facilities:} 
\facility{CXO (ACIS-I)},
\facility{Spitzer (IRAC,IRS,MIPS)}

\bibliographystyle{apj3}

\end{document}